\documentclass[12pt]{article}
\pdfoutput=1

\usepackage{cite}

\usepackage[nointlimits,reqno]{amsmath}
\usepackage{amssymb}
\usepackage{simplewick}
\usepackage{slashed}
\numberwithin{equation}{section} 

\usepackage{eucal}
\usepackage{dsfont}
\usepackage{graphicx}

\usepackage[pdftex,
 pdfproducer=TeXShop,
 pdfcreator=pdflatex]{hyperref}

\usepackage{xspace}
\usepackage{luty}

\usepackage{titlesec}

\newcommand{\zbar}{\overline{z}}
\DeclareMathOperator{\Li}{Li} 
\newcommand{\cell}[1]{\parbox[c][1.5cm][c]{1.8cm}{\centering #1}} 

\begin{document}

\begin{titlepage}

\title{Scale Anomalies, States, and Rates\\
in Conformal Field Theory}

\author{Marc Gillioz}

\address{Theoretical Particle Physics Laboratory, Institute of Physics, EPFL
\\ CH--1015 Lausanne, Switzerland}

\author{Xiaochuan Lu\  \ and\ \ Markus A. Luty}

\address{Center for Quantum Mathematics and Physics (QMAP)\\
University of California, Davis, California 95616}

\begin{abstract}

\end{abstract}
This paper presents two methods to compute scale anomaly coefficients
in conformal field theories (CFTs),
such as the $c$ anomaly in four dimensions, in terms of the CFT data.
We first use Euclidean position space to show that the anomaly coefficient
of a four-point function
can be computed in the form of an operator product expansion (OPE),
namely a weighted sum of OPE coefficients squared.
We compute the weights for scale anomalies associated with scalar operators
and show that they are not positive.
We then derive a different sum rule of the same form in Minkowski momentum space
where the weights are positive.
The positivity arises because the scale anomaly is the
coefficient of a logarithm in the momentum space four-point function.
This logarithm also determines the dispersive part,
which is a positive sum over states by the optical theorem.
The momentum space sum rule may be invalidated by UV and/or IR divergences, and we discuss
the conditions under which these singularities are absent.
We present a detailed discussion of the formalism required to compute the weights directly in Minkowski momentum space.
A number of explicit checks are performed,
including a complete example in an 8-dimensional free field theory.
\end{titlepage}

\section{Introduction}
Ken Wilson taught us to think of quantum field theory as the result of ``zooming
out'' from a local quantum theory, formalized by the renormalization group (RG).
For example, a lattice of spins can be approximated as a
continuum quantum field theory at distances much longer than the lattice spacing.
This paradigm has led to an understanding of the robustness of effective field
theories in particle physics, and universality of critical behavior in condensed
matter  physics.

The RG flow can be viewed as the effect of
a scale (dilatation) transformation.
Interacting systems generally have a nontrivial behavior under scale transformations,
leading to renormalization group flows that can display interesting phenomena
such as asymptotic freedom, dimensional transmutation, and IR fixed points.
Such nontrivial renormalization group flows are possible only
due to the fact that scale transformations are generally anomalous.
If there were no scale anomaly, the breaking of scale invariance would be simply a
matter of dimensional analysis, with all couplings being either relevant, irrelevant,
or exactly marginal.

In this paper we focus on theories at a conformally invariant fixed point
of the RG.
The IR limit of many flows may be a gapped theory or a theory of free massless
particles, but there are also many interesting theories that have a nontrivial
conformally interacting IR fixed point, such as the conformal window of QCD-like theories.
At such a fixed point, beta functions for dynamical couplings are zero
by definition, but we can have a nonzero beta function for the
self-coupling of source fields that define local operators in the theory.
For example, we can define the energy-momentum tensor by coupling the theory
to a background metric $g_{\mu\nu}$.
In 4-dimensional theories, there is a purely gravitational coupling proportional to the square of the Weyl tensor
\[
\De S = \myint d^4 x \sqrt{g} \gap \al W^{\mu\nu\rho\si} W_{\mu\nu\rho\si}
\]
that is conformally invariant.
The dimensionless coupling $\al$ is renormalized even at the fixed point,
and therefore the beta function
\[
c_T = \mu \frac{d\al}{d\mu}
\]
is a nonzero constant at the fixed point.
The coefficient $c_T$ is a basic parameter in 4-dimensional conformal field theory.
For example, it appears as the coefficient of the 2-point function of the energy-momentum
tensor, and as a coefficient of one of the 3 tensor structures in the 3-point
functions.%
\footnote{There is also an $a$ anomaly, which is not an anomaly for scale
(dilatation) transformations, but does give an anomaly for special conformal
transformations.}
It is anomalies of this kind that we will study in this paper.

The anomaly coefficients appearing in 2- and 3-point functions are trivially related to the normalization of the fields and the OPE coefficients. Starting with the 4-point function, however, the anomaly coefficients depend non-trivially on the theory.
In this paper we will study a scale anomaly in the 4-point function
of a primary operator $\scr{O}$ of the form
\[
\eql{cOdefn}
D \avg{\scr{O}(x_1) \scr{O}(x_2) \scr{O}(x_3) \scr{O}(x_4)}
= c_{\scr{O}} \de^d(x_{12}) \de^d(x_{23}) \de^d(x_{34}),
\]
where $D$ is the dilatation operator (see \Eq{thescaleanomexample} for a precise definition),
$d$ is the space-time dimension,
we define $x_{ij} = x_i - x_j$,
and $c_{\scr{O}}$ is an anomaly coefficient.
Note that such a scale anomaly can occur only if the scaling dimension of
$\scr{O}$ is given by $\De_{\scr{O}} = 3d/4$.
The motivation for studying such a special case is to develop techniques to
deal with the more interesting but more complicated cases such as the $c_T$ anomaly
in 4-dimensional CFTs.

We will show that scale anomaly coefficients such as $c_{\scr{O}}$
defined in \Eq{cOdefn} can
be written in terms of the CFT data (operator spectrum and OPE coefficients)
in an OPE-like expansion
\[
\eql{cOsumrule}
c_{\scr{O}} = \sum_\Psi \la_{\scr{OO}\Psi}^2 C_\Psi , 
\]
where the sum is over all primary operators $\Psi$ that appear in the
$\scr{OO}$ OPE, $\la_{\scr{OO}\Psi}$ are the OPE coefficients,
and $C_\Psi$ are universal constants (independent of the theory) that we
can compute.
We can think of \Eq{cOsumrule} as a manifestation of the ``infrared face''
of the scale anomaly in a CFT, since it relates the anomaly coefficient
to the properties of the CFT data, which determines the correlation functions
at separated points.

In this paper we derive two versions of the sum rule \Eq{cOsumrule}.
The first is derived in Euclidean position space.
In this case we exploit the similarity of \Eq{cOsumrule} to Gauss' law with
a point source,
$\vec\nabla \cdot \vec{E} = c_{\scr{O}} \de^{3d}(\pvec{X})$.
This can be used to express the anomaly
coefficient $c_{\scr{O}}$ in terms of a flux integral.
We give explicit expressions for $c_{\scr{O}}$ as an integral
over the conformal cross ratios
$u$ and $v$ with universal smooth weight function.
In the resulting sum rule, the
coefficients $C_\Psi$ are not positive definite.
For example, in $d = 4$ the contribution from operators of spin
$\ell = 0, 4, 8, \ldots$ is positive, while the other spins
and the identity operator give negative contributions.%
\footnote{It is somewhat misleading to think of $C_{\mathds{1}}$ as the
contribution of the identity operator because it really comes from the
complete disconnected part of the $\avg{\scr{OOOO}}$ correlator.
See the discussion around \Eq{positionC1} below.}
The second version of the sum rule is derived in Minkowski momentum space,
and gives a sum rule where the identity operator does not contribute and
$C_\Psi \ge 0$ for all operators $\Psi$.
In this case the positivity arises because the scale anomaly is related
to the coefficient of a logarithm in a forward 2-to-2 scattering amplitude,
and this same logarithm gives the imaginary part of this amplitude,
which is a positive sum over states by the optical theorem.
The scattering amplitude is given by the connected part of the
$\avg{\scr{OOOO}}$ correlation function, and therefore the identity operator
does not appear in this version of the sum rule.
This argument is phrased in the language of
scattering theory to make connections with well-known physical concepts,
but it can be translated to an argument about CFT correlators and does not require any assumption about the existence of an $S$-matrix for a CFT.

The momentum space sum rule is not without caveats.
In particular, its validity requires the absence of UV and IR divergences, which do
occur in some theories.
The UV divergences are well understood, and they restrict the validity
of the sum rule to theories where the operator spectrum satisfies some
conditions that are mild if $\scr{O}$ is a low-dimension operator.
There are also IR divergences that are less well understood.
We give some physical arguments and some examples, including a complete
test of the momentum-space sum rule for an 8-dimensional free CFT that
has no UV or IR divergences.
We leave more
detailed investigation of this question to future work.

There are a number of reasons this work may be of interest.
First, we believe that the scale anomaly is intrinsically an interesting
object of study, and we should understand all aspects of it.
At a more practical level, the sum rule for the $c$-anomaly in 4-dimensional
CFT may potentially lead to new analytic bootstrap bounds.
If we can write a positive sum rule of the form \Eq{cOsumrule} for
the $c_T$ anomaly for 4-dimensional conformal field theory, it will take the form
\[
c_T - C_T \la_{TTT}^2 = \sum_{\Psi \, \ne\, \mathds{1}, T} \la_{TT\Psi}^2 C_\Psi,
\]
where we have moved the contribution from the energy-momentum tensor
to the \lhs\ (it is actually the sum of 3 terms).
The quantity on the left is a positive-definite combination of the
3 $TTT$ OPE coefficients, and because every term in the sum is positive,
we obtain a lower bound on this quantity as a function of the OPE coefficient
$\la_{TT\scr{O}}$ for the lowest-dimension scalar operator.
The validity of this bound requires us to understand the question of
IR divergences, but it would be an example of a completely analytic
bootstrap bound that is not easily obtained from the traditional
Euclidean approach.
We leave the investigation of this sum rule for future work.

Yet another reason this work may be of interest is that we develop
the technology for summing over CFT states in Minkowski space, and in
fact in momentum space.
This is interesting because scattering amplitudes are natural observables
that live in momentum space.
We develop the analogs of conformal blocks for
4-point functions of the form
\[
\eql{WightmanOPE}
\bra{0} [\scr{O}_4 \scr{O}_3] [\scr{O}_2 \scr{O}_1] \ket{0}
= \sum_\Psi \bra{0} [\scr{O}_4 \scr{O}_3] \ket{\Psi}
\bra{\Psi} [\scr{O}_2 \scr{O}_1] \ket{0},
\]
where the factors $[\scr{O}_2 \scr{O}_1]$
and $[\scr{O}_4 \scr{O}_3]$ are ordered as shown (Wightman ordering),
while the operators within square brackets may have some
other ordering such as time ordering.
We show how to carry out such sums over states in terms of the state-operator
correspondence, so that \Eq{WightmanOPE} can be viewed as a kind of OPE.
The idea is simply to map the usual state-operator correspondence in radial
quantization in Euclidean space to Minkowski space using a combination of
conformal transformations and a Wick rotation.
Note that because the operator ordering on the two sides of the insertion
of states is fixed, the sum over states converges as long as the norm of
the states $[\scr{O}_2 \scr{O}_1]\ket{0}$
and $[\scr{O}_4 \scr{O}_3]^\dagger \ket{0}$ is finite.
This implies that we can freely Fourier transform the OPE term by
term and obtain the analogs of the conformal blocks in momentum space
as a square of Fourier transform of 3-point functions.
We carry out a number of explicit checks that verify that this sum over
states correctly reproduces known correlation functions.
The OPE for Wightman functions and its validity in momentum
space have been previously discussed in the
literature \cite{Luscher:1974ez, Dobrev:1975ru}, but the treatment is heavily mathematical
and is difficult to apply in practice.
We hope that the present more physical discussion will prove useful
for various Wightman-like correlation functions that have played a role
in recent work on conformal field theory, for example in \Refs{Hofman:2008ar, Maldacena:2015waa}.

This paper is organized as follows.
In Section 2, we derive the Euclidean position space version of the sum rule
\Eq{cOsumrule}.
In Section 3, we derive the Minkowski momentum space version of the sum rule.
Section 4 contains our conclusions.
The discussion in the main part of the paper focuses on the main
ideas, but many details are provided in the appendices.

\section{The Scale Anomaly in Euclidean Position Space}
\scl{positionspace}

We begin by discussing the scale anomaly in Euclidean position space
in arbitrary dimension $d$.
We focus on the connected 4-point function of identical scalar primary
operators $\scr{O}$ in Euclidean position space obeying the anomalous scale
Ward identity
\[
\eql{thescaleanomexample}
\left( \sum_{i = 1}^4 x_i \cdot \frac{\d}{\d x_i} + 4 \Delta_\scr{O} \right)
\avg{\scr{O}(x_1) \cdots \scr{O}(x_4)}_\text{conn}
= c_{\scr{O}} \gap \de^d(x_{12}) \de^d(x_{23}) \de^d(x_{34}).
\]
This requires the special value of the scaling dimension
\[
\De_{\scr{O}} = \frac{3d}{4}.
\]
Note that  $d = 4$ examples of this include the operator
$\scr{O} = \bar\psi\psi$ where $\psi$ is a free Dirac fermion,
and $\scr{O} = \phi^3$ where $\phi$ is a free scalar field.
The reason that \Eq{thescaleanomexample} is written for the connected
correlation function will be explained below.

\subsection{The Scale Anomaly and Flux Integral}

By translation invariance, the correlation function depends on the three
differences $x_{12}, x_{23}, x_{34}$, so we write
\[
\avg{\scr{O}(x_1) \cdots \scr{O}(x_4)}_\text{conn} = F(\!\ggap \vec{X}),
\qquad
\vec{X} = (x_{12}, x_{23}, x_{34}) \in \mathds{R}^{3d}.
\]
\Eq{thescaleanomexample} can then be written as
\[
\eql{Gausslaw}
\vec{\nabla} \cdot \vec{E}(\!\ggap \vec{X}) = c_{\scr{O}} \gap \de^{3d}(\!\ggap \vec{X}),
\]
where
\[
\eql{Edefn}
\vec{E} = \vec{X} F(\!\ggap \vec{X}).
\]
\Eq{Gausslaw} has the form of Gauss' law for the electric field in the presence of
a point charge $c_{\scr{O}}$ at $\vec{X} = 0$.
The electric field away from $\vec{X} = 0$ is given by the CFT data, and our
job is to find the value of $c_\scr{O}$ corresponding to that.
The solution is to write $c_{\scr{O}}$ as a flux integral
\[
\eql{cfluxint}
c_{\scr{O}} = \int_{\Si} d\vec{\Si} \cdot \vec{E},
\]
where $\Si$ is any codimension-1 surface in $\mathds{R}^{3d}$ that encloses
$\vec{X} = 0$.

The integral in \Eq{cfluxint} depends on the correlation function at separated
points, and shows  that the anomaly coefficient $c_{\scr{O}}$ can be
computed by the CFT data.
(The integral includes points $\vec{X}$ where 2 or 3 points coincide,
but we will see below that these singularities are integrable.)
We can interpret \Eq{thescaleanomexample} as saying that the divergence of the correlation
function vanishes everywhere except at the coincident points, where it has a delta
function singularity.
Just as in electrostatics, the behavior of the fields at finite distances determines
the strength of the delta function singularity, explaining why it is independent of
any short-distance regulator.

We now show how to compute the flux integral \Eq{cfluxint} in terms
of the CFT data.
Writing the 4-point function using the $(12)(34)$ OPE, we have
\[
\eql{confblockexpansion}
\avg{\scr{O}(x_1) \cdots \scr{O}(x_4)}
= \frac{1}{(x_{12}^2 x_{34}^2)^{\De_{\scr{O}}}}
\!\!\ \sum_{\Psi \,=\, \text{primary}} \!\!
\la^2_{\scr{O}\scr{O}\Psi} g_{\Psi}(u, v),
\]
where the sum over $\Psi$ runs over all primary operators,
$\la_{\scr{O}\scr{O}\Psi}$ is an OPE coefficient,
and $g_{\Psi}$ is a conformal block
depending on the conformally invariant cross-ratios
\[
\eql{uvcrossratios}
u = \frac{x_{12}^2 x_{34}^2}{x_{13}^2 x_{24}^2},
\qquad
v = \frac{x_{14}^2 x_{23}^2}{x_{13}^2 x_{24}^2}.
\]
The OPE expansion is valid only
for the full 4-point function, including the disconnected part:
\beq
\begin{split}
\avg{\scr{O}(x_1) \cdots \scr{O}(x_4)}
&= \avg{\scr{O}(x_1) \cdots \scr{O}(x_4)}_\text{conn}
\\
&\qquad{}
+ \frac{1}{x_{12}^{2\De_{\scr{O}}} x_{34}^{2\De_{\scr{O}}}}
+ \frac{1}{x_{13}^{2\De_{\scr{O}}} x_{24}^{2\De_{\scr{O}}}}
+ \frac{1}{x_{14}^{2\De_{\scr{O}}} x_{23}^{2\De_{\scr{O}}}}. 
\end{split}
\eeq
The disconnected part can have its own scale anomaly.
For $d = 4k$ with integer $k$, we have
\[
\left( \sum_{i = 1}^4 x_i \cdot \frac{\d}{\d x_i} + 4 \Delta_\scr{O} \right)
\frac{1}{x_{12}^{2\De_{\scr{O}}} x_{34}^{2\De_{\scr{O}}}}
= & \left[ \left( x_1 \cdot \frac{\d}{\d x_1} + x_2 \cdot \frac{\d}{\d x_2} + 2 \Delta_\scr{O}
\right)
\frac{1}{x_{12}^{2\De_{\scr{O}}}} \right]
\frac{1}{x_{34}^{2\De_{\scr{O}}}}
\nn
&
+ \frac{1}{x_{12}^{2\De_{\scr{O}}}}
\left( x_3 \cdot \frac{\d}{\d x_3} + x_4 \cdot \frac{\d}{\d x_4} + 2 \Delta_\scr{O}
\right)
\frac{1}{x_{34}^{2\De_{\scr{O}}}}
\nonumber\\[4pt]
\sim \, & \d^{2k} \de^d(x_{12}) \frac{1}{x_{34}^{2\De_{\scr{O}}}}
+ \frac{1}{x_{12}^{2\De_{\scr{O}}}} \d^{2k} \gap \de^d(x_{34}).
\]
This gives a contribution to the anomaly that
involves products of local and non-local terms.
On the other hand, the anomaly for the connected correlation function is completely local,
in the sense that the \rhs\ of \Eq{thescaleanomexample} is a delta function
where all points are coincident.
This is a manifestation of the fact that the scale anomaly for the quantum
effective action is local:
\[
\de_\si W = \myint d^d x\ggap c_{\scr{O}}  \si \gap \frac{\rho_{\scr{O}}^4}{4!},
\]
where $\rho_{\scr{O}}$ is the source field for the operator $\scr{O}$.
The locality of the transformation of the quantum effective action is the defining
property of an anomaly.
It arises from the fact that the anomaly can be viewed as an effect of the
UV regulator, which implies that the connected correlation functions change
only by local terms under scale transformations.

We therefore obtain a sum rule of the form
\[
\eql{Euclideansum}
c_{\scr{O}} = C_\text{disc} + \sum_{\Psi}
 \la_{\scr{O}\scr{O}\Psi}^2 C_\Psi,
\]
where $C_\text{disc}$ is the contribution to the flux integral
from the disconnected part
\[
C_\text{disc} = -\int_{\Si} d\vec{\Si} \cdot \vec{E}_\text{disc}.
\]
where $\vec{E}_\text{disc}$ is the contribution to $\vec{E}$ from
the disconnected part of the 4-point function.
Because the 2-point function is non-negative in Euclidean space,
we have $C_\text{disc} \le 0$.

\subsection{Computing the Flux Integral}
We now present the main ideas and results of the computation of the flux integral
\Eq{cfluxint}.
Details are given in Appendix~\ref{sec:fluxintegral}.

The flux integral is over a $3d - 1$ dimensional space, but the
$\avg{\scr{OOOO}}$ correlation function that gives the integrand depends
non-trivially only on the cross ratios $u$ and $v$.
It is therefore clear that we can rewrite the flux integral as an
integral of conformal blocks $g_\Psi(u, v)$ over $u$ and $v$ weighted
by a universal weight function that depends only on $\De_{\scr{O}}$.
The strategy to obtain this representation of the flux integral is to use
the Faddeev-Popov method to factor out the remaining integrals.
The result is
\[
\begin{split}
\eql{c_position_space}
c_{\scr{O}} &= \int\limits_{\Im(z) > 0}
	\frac{d z d \zbar}{(\Im z)^2} \,
	\scr{K}_d(u, v)
	\biggl[ \left( \frac{v}{u^2} \right)^{d/4}
	\sum_\Psi \lambda_{\scr{O}\scr{O}\Psi}^2
	g_\Psi (u, v)
\\
&\qquad\qquad\qquad\qquad\qquad\quad{}
	- \left( \frac{v}{u^2} \right)^{d/4}
	- \left( u v \right)^{d/4}
	- \left( \frac{u}{v^2} \right)^{d/4} \biggr],
\end{split}
\]
in terms of a complex variable $z$ and its conjugate $\zbar$.
The term in square brackets is proportional to the connected part of the 4-point function:
it is the difference between the conformal block expansion of \Eq{confblockexpansion} and
the disconnected part.
This term is invariant under crossing
$(u, v) \leftrightarrow (v,u) \leftrightarrow (1/u, v/u)$.
(Note that we are using the conventional definition of $g_\Psi(u, v)$ which has a
nontrivial transformation under crossing, see \Eq{confblockexpansion}.)
The variables $z$ and $\zbar$ in \Eq{c_position_space} are the natural
variables for the conformal blocks \cite{Dolan:2000ut, Dolan:2003hv},
defined by
\[
	\eql{z_definition}
	u = |z|^2,
	\qquad
	v = |1 - z|^2.
\]
The integration kernel $\scr{K}_d$ is given by
\[
	\eql{int_kernel}
	\scr{K}_d ( z, \zbar )
	& = \frac{4 \, \pi^{(3d-1)/2}}
	{\Gamma\left(\frac{d-1}{2} \right)
	\Gamma\left(\frac{d}{2} \right)
	\Gamma\left(\frac{d}{4} \right)^2} \,
	\frac{|\Im z|^d}{|z|^{d/2} |1-z|^{d/2}}
	\nonumber \\
	&
	\qquad\qquad{} \times
	\int_0^1 d\lambda_1 d\lambda_2 d\lambda_3
	\frac{\delta(\lambda_1 + \lambda_2 + \lambda_3 - 1)
	(\lambda_1 \lambda_2 \lambda_3)^{d/4-1}}
	{(\lambda_1 \lambda_2 + u \lambda_1 \lambda_3
	+ v \lambda_2 \lambda_3)^{d/4}}.
\]
In these variables, crossing symmetry follows from invariance under
$z \leftrightarrow 1-z$ and $z \leftrightarrow z^{-1}$.
This can be used to check that the integration measure in \Eq{c_position_space}
as well as the integral \eq{int_kernel} are invariant under crossing.
The kernel $\scr{K}_d$ is a positive function with a maximum at the crossing symmetric
point $u = v = 1$, and is analytic everywhere except $z = 0$ and $z = 1$.
A plot of $\scr{K}_4$ is given in \Fig{K4}.
More details on the structure of the kernel are presented in Appendix~\ref{sec:fluxintegral}.
\begin{figure}
	\centering
	\includegraphics[scale=0.6]{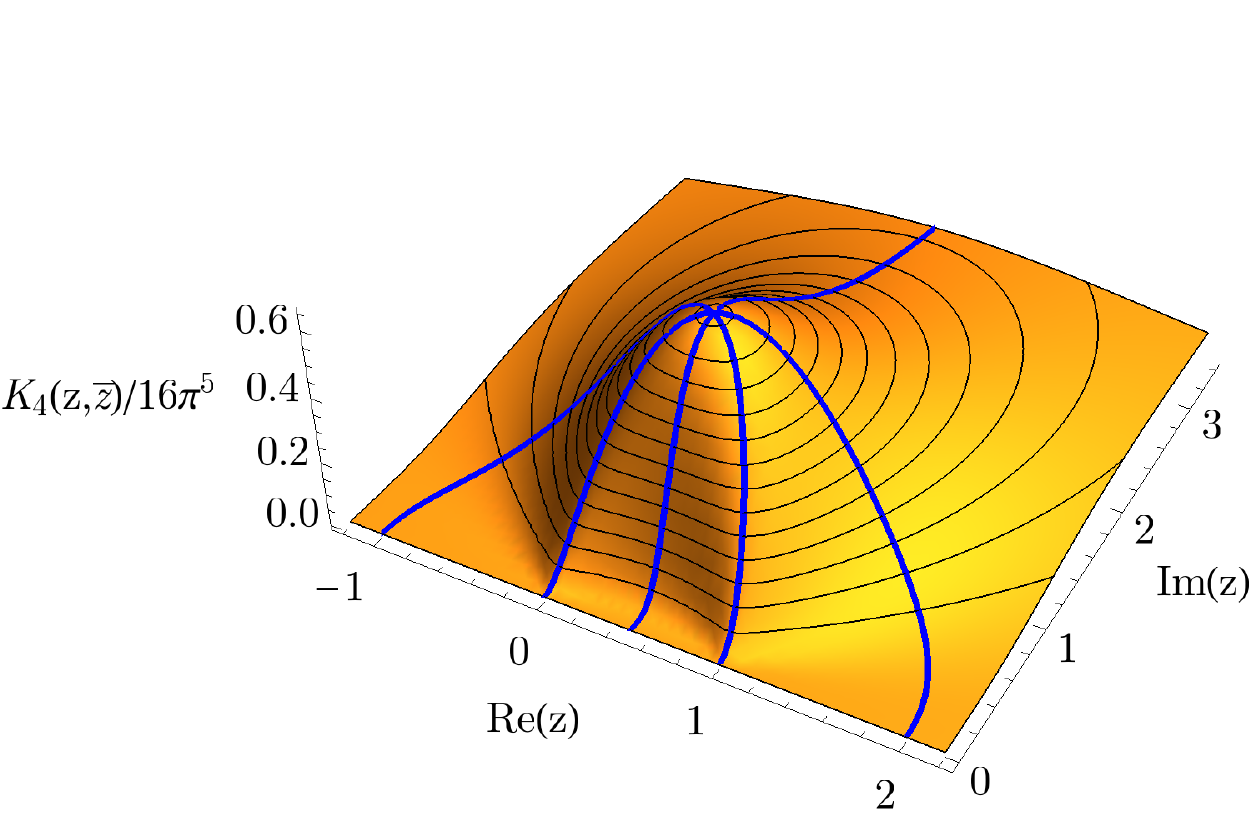}
	\caption{The integration kernel function $\scr{K}_4$ in $d=4$ in the complex $z$ plane. The blue	lines separate regions related by crossing-symmetry. The maximum of $\scr{K}_4$ occurs at the completely crossing-symmetric point $u = v = 1$.}
	\label{fig:K4}
\end{figure}%

The integral \Eq{c_position_space} goes over the entire upper half complex $z$ plane,
which includes regions where the OPE in \Eq{confblockexpansion}
does not converge.
However, we can use different OPE channels in different regions of integration
so that only convergent OPE expansions are used.
Since regions with different convergent OPEs are related to each other by crossing
symmetry, we can equivalently restrict the domain of integration to a fundamental domain.
One example is
\[
	\eql{fundamentaldomain}
	0 < u < v < 1
	\qquad \Leftrightarrow \qquad
	|1-z|^2 < 1,\  \Re(z) < 1/2,\ \Im(z) > 0.
\]
\begin{figure}
	\centering
	\includegraphics[width=6.8cm]{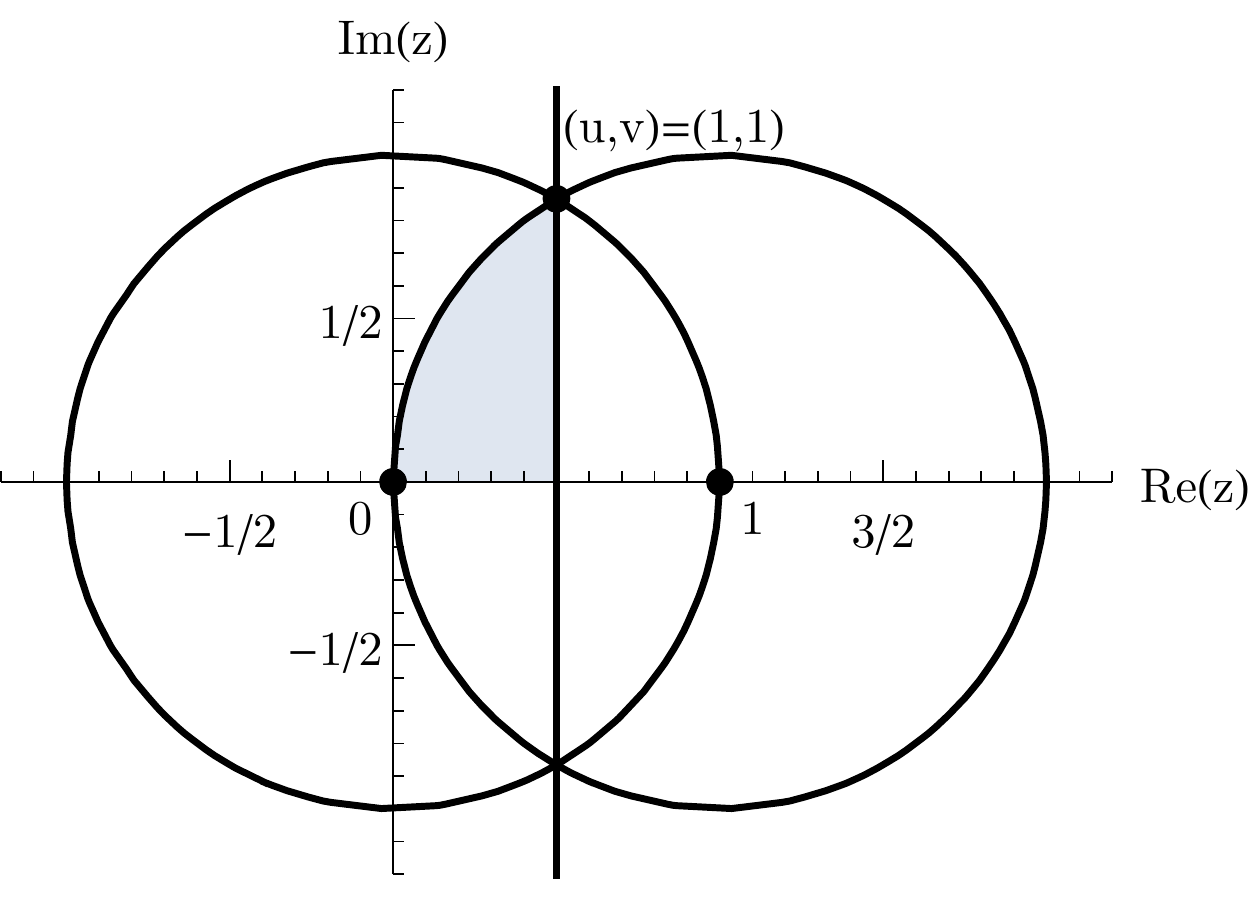}
	\caption{The fundamental domain of integration defined by the constraints of \Eq{fundamentaldomain}. Crossing symmetry corresponds to reflections across the line at $\Re(z)= \frac{1}{2}$ and the two circles centered at $z=0$ and $z=1$. }
	\label{fig:fundamentaldomain}
\end{figure}%
This domain is illustrated in \Fig{fundamentaldomain}.
In this case we are integrating only over regions where the OPE converges.

When this is done, we find that the integral \Eq{c_position_space}
converges everywhere except possibly at $z = 0$, where the kernel has the
limiting behavior
\[
\eql{Katzero}
\scr{K}_d(z) \sim |z|^{d/2} \log|z|.
\]
Convergence of the integral near $z = 0$ requires therefore
\[
\De_\Psi > \frac{d}{2}.
\]
For $d>2$, the unitarity constraints imply that divergences can only happen for scalar operators,
and we conclude that our sum rule \Eq{Euclideansum} is invalid for theories with
``super-relevant'' scalar operators $\Psi$ in the $\scr{OO}$ OPE
($\frac{d-2}{2} \le \De_\Psi \le \frac{d}{2}$).
The integral for $\Psi = \mathds{1}$ also diverges, but it is canceled by the contribution
from part of the disconnected piece.
We can therefore rewrite \Eq{Euclideansum} as
\[
\eql{Euclideansum:2}
c_{\scr{O}} = C_{\mathds{1}} + \sum_{\Psi \, \ne \, \mathds{1}} \la_{\scr{OO}\Psi}^2 C_\Psi,
\]
where
\[
\eql{positionC1}
C_{\mathds{1}} &= -6 \int\limits_{0 < u < v < 1}
\frac{dz d\zbar}{(\Im z)^2}\,
\scr{K}_d(z, \zbar)
\left[ (uv)^{d/4} + \left( \frac{u}{v^2} \right)^{d/4} \right],
\\
C_\Psi &= 6 \int\limits_{0 < u < v < 1}
\frac{dz d\zbar}{(\Im z)^2}\,
\scr{K}_d(z, \zbar)
\left( \frac{v}{u^2} \right)^{d/4} g_\Psi(u, v),
\]
where the factors of 6 account for the fact that there are 6 fundamental domains.
Associating $C_{\mathds{1}}$ with the unit operator is a slight abuse of notation, since
we see that it is really associated with the portion of the disconnected contribution
that is \emph{not} the unit operator.

\begin{figure}
	\centering
	\includegraphics[scale=0.8]{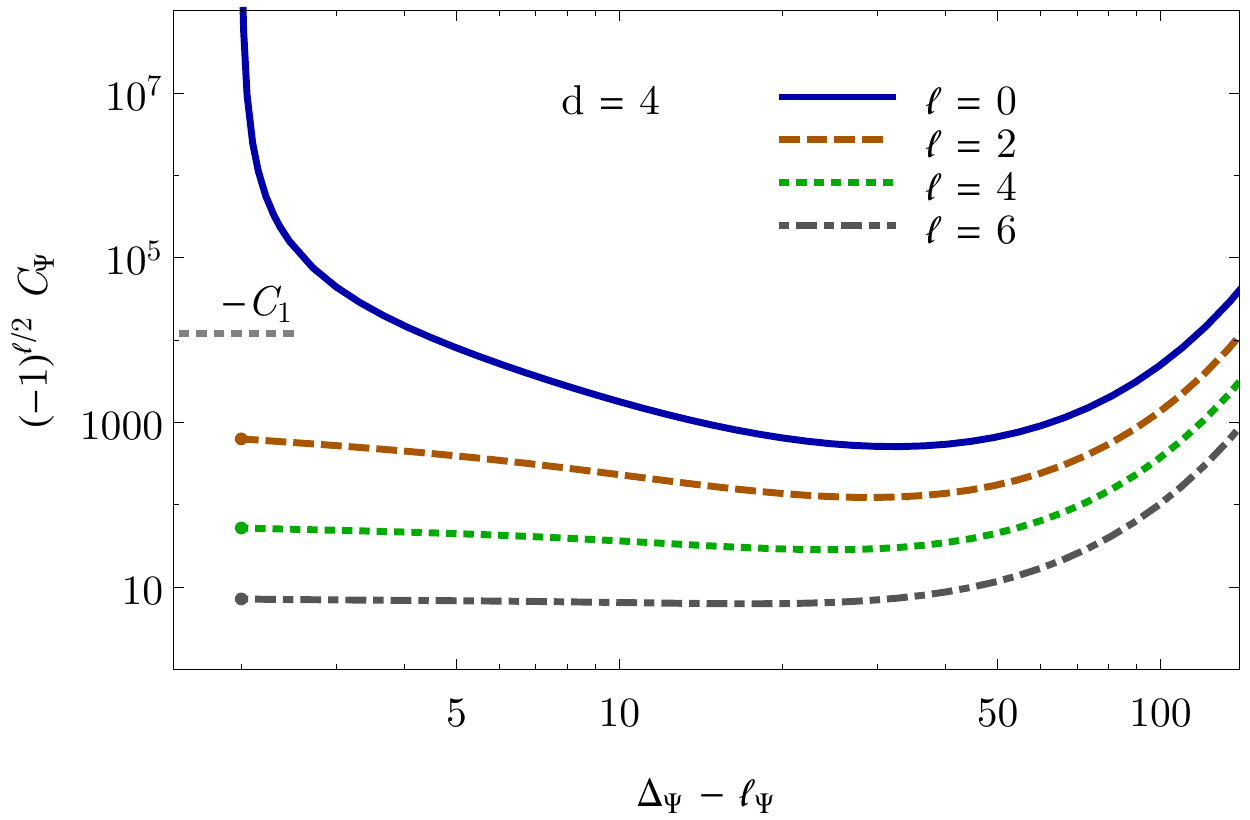}
	\caption{Numerical value of the coefficient $C_\Psi$ of \Eq{Euclideansum:2} in $d = 4$ as a function of the scaling dimension $\Delta_\Psi$ and spin $\ell_\Psi$ of the intermediate operator $\Psi$. The value of the universal contribution $C_\mathds{1}$ is shown as well for reference.}
	\label{fig:position_space_coeffs}
\end{figure}

It is interesting to know the signs of the coefficients $C_\Psi$.
A numerical analysis of the $d = 4$ case shows an alternating pattern:
for operators $\Psi$ with spin $\ell = 2n$,
those with even (odd) $n$ give a positive (negative) contribution, while $C_\mathds{1} < 0$.
It is somewhat disappointing that the sum does not have a definite sign, but we will derive
a positive-definite sum rule in \sec{momentumspace}.
The series does appear to converge very quickly.
One piece of evidence for this comes from the values of the coefficients $C_\Psi$ themselves,
which are shown in \Fig{position_space_coeffs}.
Note that these decrease rapidly as a function of the twist $\tau = \De - \ell$
up to about $\tau \sim 30$.
For $\tau \gsim 30$, the coefficients increase exponentially, but the OPE coefficients
are expected to decrease exponentially as well~\cite{Pappadopulo:2012jk, Fitzpatrick:2012yx, Rychkov:2015lca}.
We also note that at fixed twist, the coefficient for scalar operators always dominates
by a large factor.
We leave more detailed investigation of the convergence of this sum rule to future work.

\subsection{An Example: the Free Scalar in 8 Dimensions}

The free scalar field theory in $d=8$ dimensions provides an interesting test of our sum rule: the free field $\phi$ having scaling dimension 3, there is an anomaly of the form discussed above for the operator
$\scr{O} = \phi^2 / \sqrt{2}$,
where the normalization is chosen so that the
2-point function takes the canonical form.
Since the theory is free, we also know directly the 4-point function in terms of the conformal cross-ratios $u$ and $v$, see \Eq{free4ptfct} of the appendix.
Extracting the connected part, we obtain an expression for the anomaly as a flux integral,
\[
	\eql{freescalaranomaly}
	c_{\phi^2} = 6 \int\limits_{0 < u < v < 1}
	\frac{d z d \zbar}{\left( \Im{z} \right)^2} \,
	\scr{K}_8 \left( z, \zbar \right)
	\left[ \frac{4}{u v} + \frac{4 u^2}{v}
	+ \frac{4 v^2}{u} \right]
	= \frac{\pi^{12}}{4}.
\]
where $\scr{K}_8$ is provided explicitly in \Eq{int_kernel_8d}.

We have not performed the sum over conformal blocks in this case, since the calculation of
the OPE is very cumbersome.
However, we can test the numerical convergence by computing the contribution of low-lying
operators.
The results are shown in Table~\ref{tab:scalar_d8_example}.
We see that there is good evidence that the sum is converging rapidly.
\begin{table}
	\centering
	\begin{tabular}{|c|c|ccc|cc|}
		\hline
		$\Psi$ &
		$\mathds{1}$ &
		\cell{$\frac{1}{\sqrt{2}} \phi^2$} &
		\cell{$T^{\mu\nu}$} &
		\cell{$\partial^n \phi^2$ \\ $(n \geq 4)$} &
		\cell{$\frac{1}{2\sqrt{6}} \phi^4$} &
		\cell{$\partial^n \phi^4$ \\ $(n \geq 2)$}
		\\ \hline
		\parbox[c][1cm]{1cm}{\centering $\Delta_\Psi$} &
		0 & 6 & 8 & $\geq 10$ & 12 & $\geq 14$
		\\ \hline
		\cell{$\dfrac{\lambda_{\phi^2\phi^2\Psi}^2 C_\Psi}{c_{\phi^2}}$} &
		$-0.03$ &
		$1.02$ &
		$-0.05$ &
		$0.01$ &
		$0.08$ &
		$-0.02$
		\\ \hline
	\end{tabular}
	\caption{The contribution of all the operators in the $\phi^2\phi^2$ OPE to the scale anomaly $c_{\phi^2}$ of \Eq{freescalaranomaly}. The second column indicates the disconnected part. The operator $\phi^2$ itself gives by far the largest contribution to the anomaly. $T^{\mu\nu}$ is the stress-energy tensor for the free scalar theory. The columns denoted $\partial^n \phi^2$ and $\partial^n \phi^4$ indicate the total contribution of all operators of each type.}
	\label{tab:scalar_d8_example}
\end{table}

\section{The Scale Anomaly and Rates in Minkowski Momentum Space}
\scl{momentumspace}

We now give a different sum rule for the scale anomaly by relating
it to a property of physical observables.

\subsection{Scale Anomalies and Rates}
The idea is to couple the operator $\scr{O}$ to a free quantum
probe field $A$ via
\[
\eql{SintA}
\De S = \myint d^d x\, \ep \gap A \gap \scr{O},
\]
where $\ep$ is a coupling constant with dimension $d - \De_{\scr{O}}$.
We work to lowest order in $\ep$ for processes involving external $A$
particles (no $A$ loops), so that we can view $A$ as a probe of the
CFT dynamics.
This is analogous to using the photon to probe QCD.

Because we want to study the 4-point function of $\scr{O}$, we
consider the $AA \to AA$ scattering amplitude at leading order
in $\ep$:
\[
i\scr{M}(AA \to AA) =
\frac{\ep^4}{V} \bra{0} T[ \scr{O}(p_1) \cdots \scr{O}(p_4)] \ket{0},
\]
where we define the Fourier transformed operators
\[
\scr{O}(p) = \myint d^d x \gap e^{i p \cdot x} \scr{O}(x),
\]
and division by the space-time volume $V = (2\pi)^d \de^d(p = 0)$ cancels the overall
momentum-conserving delta function.
Note that the disconnected contribution has two momentum-conserving
delta functions, and vanishes for general momenta.
We will later consider forward kinematics, but it should be understood as a limit,
so that the disconnected part never contributes to the amplitude.

We take $A$ to be massless ($p_i^2 = 0$), so the amplitude is a function of the usual
Mandelstam invariants
\beq
\begin{split}
s &= -2 \, p_1 \cdot p_2 = -2 \, p_3 \cdot p_4,
\\
t &= -2 \, p_1 \cdot p_3 = -2 \, p_2 \cdot p_4,
\\
u &= -2 \, p_1 \cdot p_4 = -2 \, p_2 \cdot p_3,
\end{split}
\eeq
with
\[
s + t + u = 0.
\]
Note that we define all momenta to be in-going.
In the forward limit $t \to 0$ the amplitude is a function only of $s$.
(In the following we assume that the $t \to 0$ limit is well-defined.
This point will be discussed further below.)
Crossing symmetry $s \leftrightarrow u$ implies that $\scr{M}(-s) = \scr{M}(s)$,
and scale invariance then fixes the amplitude to be
\[
\eql{Mnaive}
\scr{M}(s) = \al \ep^4 s^{2\De_{\scr{O}} - 3d/2} + (s \leftrightarrow -s).
\]
Real analyticity of scattering amplitudes
$\scr{M}^*(s) = \scr{M}(s^*)$
tells us that the constant $\al$ is real.
For general $\De_{\scr{O}}$, the amplitude has a cut along the real
$s$ axis, and we have
\[
\Im\scr{M}(s) = -\al \ep^4 s^{2\De_{\scr{O}} - 3d/2}
\ggap \sin \bigl[ \pi(2\De_{\scr{O}} - 3d/2) \bigr].
\]
Note that $\Im\scr{M}$ vanishes whenever
\[
2\De_{\scr{O}} - 3d/2 = r = 0, 1, 2, \ldots.
\]
This cannot be right, since by the optical theorem the imaginary part of the
forward amplitude is proportional to the total annihilation cross section
$AA \to \big( \text{CFT states} \big)$, which is surely nonzero.

Note however that this vanishing occurs precisely where the special
values of the dimensions allows a scale anomaly of the form
\[
\left( \sum_{i=1}^4 x_i \cdot \frac{\d}{\d x_i}
+ 4 \De_{\scr{O}} \right)
\avg{T\left[\scr{O}(x_1) \cdots \scr{O}(x_4)\right]}_\text{conn}
\sim c_{\scr{O}}^{(r)} \ggap \d^{2r} \bigl[ \de^d(x_{12})
\de^d(x_{23}) \de^d(x_{34}) \bigr].
\eql{scaleanomalyCFTscalarexample}
\]
This expression is schematic because for $r > 0$ there may be more
than one structure on the \rhs.
The scale anomaly modifies \Eq{Mnaive} to
\[
\scr{M}(s) = \ep^4 \left[ \al s^r  + \be s^r \ln s \right]
+ (s \leftrightarrow -s).
\]
The log term arises from \Eq{scaleanomalyCFTscalarexample},
and so $\be$ is a linear combination of the
scale anomaly coefficients $\sim c_{\scr{O}}^{(r)}$.
This gives for the imaginary part
\[
\Im \scr{M}(s) = -\pi \be \ep^4 s^r.
\]
We see that the imaginary part of the amplitude is determined by scale anomalies
in all cases where they are present.%
\footnote{If the dimension $\De_{\scr{O}}$ continuously approaches a value
where there is a scale anomaly, we expect that $\mathcal{M}(s) \sim 1/\de$ where
$\de = 2\De_{\scr{O}} - 3d/2 - r$.
Then as $\de \to 0$ the imaginary part of the amplitude
is finite,
and the real part is local and divergent,
but this can be canceled by adding a local counterterm $\sim \d^{2r} \rho^4 / \de$
to the theory to make the amplitude finite.
In this way, everything depends smoothly on $\De_{\scr{O}}$.}

We will later restrict the discussion to the case of the $r = 0$ anomaly
defined in \Eq{cOdefn}.
In that case we have $\beta = -c_\scr{O}/4$
and we learn that $c_{\scr{O}} > 0$.
In fact, $c_{\scr{O}}$ can be written as a positive sum over states.
In a CFT, the states are in one-to-one correspondence with the operators
of the theory, so we expect a sum rule of the form
\[
\eql{csumrule}
c_{\scr{O}} = \sum_{\Psi \, \ne \, \mathds{1}} \la_{\scr{O}\scr{O}\Psi}^2 C_\Psi,
\]
with $C_\Psi \ge 0$.
Note that the identity operator does not appear because it is purely
disconnected, and does not contribute to scattering.
Our main goal in this section is to make this sum rule precise
and compute the coefficients $C_\Psi$.
In the derivation of this rule, however, we will mostly make statements about the imaginary part of the amplitude, i.e.~we will work in the general case where $\Delta_\scr{O}$ is not necessarily equal to $3d/4$, and take the special case corresponding to the anomaly at the end.

The above discussion has used the language of scattering theory
because it relates the argument to well-known physical concepts.
However, it is important to realize that our results do not depend on the assumption
that the theory where $A$ is coupled to the CFT has a well-defined $S$ matrix.
The ``amplitude'' $\scr{M}$ is defined by the Fourier transform
of the 4-point function $\avg{T \left[ \scr{O} \scr{O} \scr{O} \scr{O} \right]}$.
The ``optical theorem'' follows from the combinatoric operator identity
\[
\eql{algebraicoptical}
\sum_{k \, = \, 0}^n (-1)^k \!\!\!\!
\sum_{\si \, \in \, \Pi(k, n - k)} \!\!
\widebar{T}[ \scr{O}(x_{\si_1}) \cdots \scr{O}(x_{\si_k})]
T[ \scr{O}(x_{\si_{k+1}}) \cdots \scr{O}(x_{\si_n})] = 0,
\]
where the sum over $\si$ is over all partitions of $1, \ldots, n$ into two groups,
and $T$ ($\widebar{T}$) denotes time ordering (anti-time ordering).
This result is proved by writing out all the (anti-)time orderings and checking
that they cancel pairwise.
For the case $n = 4$ and the region of momentum space we are assuming,
the only nonzero terms give the optical theorem that we will use to compute
the coefficients $C_\Psi$ in \Eq{csumrule}.

Before we turn to computation of these coefficients,
we address the possibility of UV and IR divergences that potentially invalidate
the result.

\subsection{UV Divergences}
\scl{UVdivergences}

The amplitude $\scr{M}(AA \to AA)$ may be ill-defined due to UV divergences.
A UV divergence that corresponds to a local counterterm for $AA \to AA$
does not contribute to
$\Im\scr{M}$, and therefore does not affect the argument above.
On the other hand, $\Im\scr{M}$ may have a UV divergence corresponding
to a counterterm for the process $AA \to \text{CFT}$ of the form
\[
\eql{UVdivergencescalarprimary}
\De S = \myint d^d x \gap \La^n \ep^2 A^2 \Psi,
\qquad n \ge 0,
\]
where $\Psi$ is a scalar primary operator in the $\scr{O} \scr{O}$ OPE, and
$\La$ is a UV cutoff mass scale.
If such a UV divergence is present, it means that the amplitude
$AA \to \text{CFT}$ depends on the dimensionful cutoff $\La$,
invalidating the arguments above.
We can remove the dependence on $\La$ by adding a counterterm to the theory,
but then the finite part of $\Im\scr{M}$ is not uniquely determined by
the coupling of \Eq{SintA}.
Dimensional analysis tells us that there can be no UV divergence
of the form \eq{UVdivergencescalarprimary} for
\[
\eql{UVdivergencebound}
\De_{\Psi} > 2 \De_{\scr{O}} - d,
\]
or equivalently $\Delta_\Psi > d/2$ in the presence of the $r=0$ anomaly.
Whenever there is no such low-dimension operator $\Psi$, we cannot have
UV divergences, and the rate is calculable in terms of $\ep$.

There are also possible counterterms involving derivatives, which can all
be written in the form
\[
\eql{UVdivergenvectorprimary}
\De S = \myint d^d x \gap \La^m \ep^2 A^2 \d_{\mu_1} \cdots \d_{\mu_\ell}
\Psi_\ell^{\mu_1 \cdots \mu_\ell},
\]
where $\Psi_\ell^{\mu_1 \cdots \mu_\ell}$ is a primary operator with spin $\ell \ge 1$
that appears in the $\scr{O}\scr{O}$ OPE.
(Note that we can use integration by parts to move all derivatives to
act on the CFT operator.)
The condition that there is no UV divergence of this kind can be written
\[
\De_{\Psi_\ell} > 2\De_{\scr{O}} - d - \ell .
\]
The conditions that this imposes amount to the statement that the operator
$\scr{O}$ have small dimension, and that operators that appear in the
$\scr{O}\scr{O}$ OPE have sufficiently large dimension.
For example, there are no UV divergences in theories where $\scr{O}$ is the only
relevant scalar operator.

\begin{figure}
	\centering
	\includegraphics[scale=1]{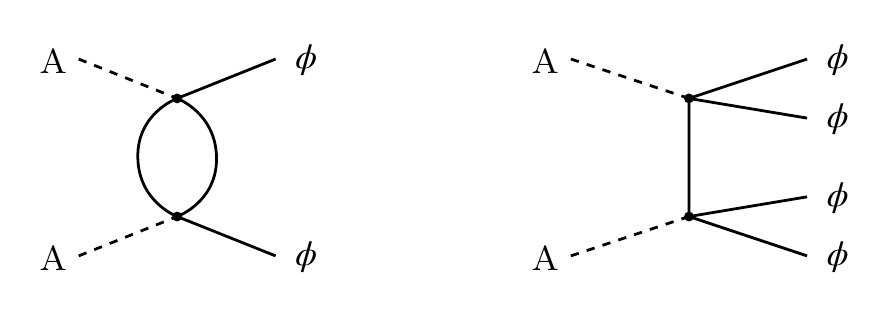}
	\caption{The two types of diagrams that contributes to the rate
	$AA \to \text{CFT}$ in the free scalar theory with $\scr{O} = \phi^2$,
	with final states
	corresponding to the primary operator $\phi^2$ (left),
	$\phi^4$ (right), or their descendants.}
	\label{fig:UVdivergence}
\end{figure}

UV divergences of the kind we are discussing are present in simple models.
For example, consider the free scalar field theory with $\scr{O} = \phi^3$.
The rate for $AA \to \text{CFT}$ is given by the two classes of diagrams shown
in \Fig{UVdivergence},
and the loop diagram is UV divergent for $d \ge 4$.
This coincides precisely with the case where the operator $\Psi = \phi^2$ violates
the bound \Eq{UVdivergencebound}.

\subsection{IR Divergences}
\scl{IRdivergences}

We now consider the possibility of IR divergences.
The most obvious worry is that the rate
$AA \to \text{CFT}$ is infinite.
In a free CFT, this rate can be computed by Feynman diagrams,
and involves an integral over final states that can be IR divergent.
One example is the free scalar theory with $\scr{O} = \phi^2$.
In this theory the rate for $AA \to \text{CFT}$
is divergent for $d \le 6$ because the integral over final states is divergent
when the $\phi$ particles have forward kinematics.
Another example is the free fermion theory with $\scr{O} = \bar\psi \psi$,
where the rate is similarly IR divergent for $d \le 4$.
This divergence can be understood as a consequence of the existence of
long-range interactions in these theories.
In a general CFT where we do not have particle states, we may expect the
same divergence in  $\Im\scr{M}(AA \to AA)$ in the limit $t \to 0$.

In the free-field theory examples we have considered,
the IR divergence occurs precisely in the case
where the coupling  $\ep$ in \Eq{SintA} is relevant or marginal.
When $\ep$ is irrelevant,
the interaction \eq{SintA} becomes weaker at small energies,
and we expect that we can compute observables in a simple series expansion
in powers of $\ep$.
On the other hand, if $\ep$ is marginal or irrelevant, we may need to resum
the series, and we may expect IR divergences if we work at fixed order in $\ep$.
If this physical argument is correct, we expect no IR divergences in the imaginary part of the amplitude when
\[
	\De_{\scr{O}} > \frac{d + 2}{2},
\]
or when considering the scale anomaly $c_\scr{O}$ with $\Delta_\scr{O} = 3d/4$, no IR divergences in dimensions $d > 4$.
Note that this can be compatible with the absence of UV divergences,
\Eq{UVdivergencebound}.
In fact, we will see that there are known theories where both UV and IR
divergences are absent, so our sum rule has some domain of applicability.

There is a more subtle type of IR divergence that can spoil the connection
between the scale anomaly and the imaginary part of the amplitude.
If the amplitude $\scr{M}(AA \to AA)$ has a finite limit as $t \to 0$,
we have seen that the scale anomaly $c_\scr{O}$ is related to the imaginary part of the amplitude:
the dilatation operator acting on the amplitude is
\[
	D \scr{M} = 2 s \frac{\d}{\d s} \scr{M} = 4 \be \epsilon^4.
\]
This gives the relation between the scale anomaly and the rate presented above.
However, imagine that for $t \ne 0$ the amplitude has the form
\[
\scr{M}(AA \to AA) = \epsilon^4 \{\al + \be \left[
\ln(-s) + \ln(-t) + \ln(-u) \right] \}.
\]
Note that this amplitude is manifestly crossing symmetric,
and is real analytic if $\al$ and $\be$ are real.
The real part of $\scr{M}$ is singular as $t \to 0$, but the imaginary
part is finite:
\[
\lim_{t \to 0_-} \Im\scr{M} = -\pi\be \epsilon^4.
\]
On the other hand, the dilatation operator acting on this amplitude for $t \ne 0$
is given by
\[
	D \scr{M}
	= 2 \left(s \frac{\d}{\d s} +
t \frac{\d}{\d t} + u \frac{\d}{\d u} \right) \scr{M} = 6 \be \epsilon^4.
\]
The relationship between the scale anomaly and the imaginary
part of the amplitude is different from earlier, due to the divergence in the real
part of the amplitude as $t \to 0$.
We see that an IR divergence in the real part of $\scr{M}$ can invalidate the
relation between the imaginary part of the amplitude and the scale anomaly,
even if both are IR finite.

We would like to have a rigorous understanding of IR divergences.
We will see below that IR divergences we are discussing do not show up as
singularities in the individual coefficients $C_\Psi$ in our sum rule
\Eq{csumrule}.
This means that the singularity as $t \to 0$ can only be due to a failure of
the sum over operators to converge.
This makes sense, because the divergences are arising from the forward limit
$t \to 0$.
Since this is localized in angle, it involves states of high angular momentum,
which come from operators with large spin.
We can hope to get a rigorous conditions that rule out the existence of IR
divergences by using bounds on the growth of OPE coefficients for large
dimensions \cite{Pappadopulo:2012jk, Fitzpatrick:2012yx, Rychkov:2015lca},
but we leave this for future work.

In this paper, we will proceed without a rigorous understanding of IR
divergences, and the reader should be aware that the absence of such
divergences is a non-trivial assumption required for the validity of \Eq{csumrule}.

\subsection{States in Minkowski Space}
Using the optical theorem as given in \Eq{algebraicoptical} above, the imaginary
part of the forward amplitude $\scr{M}$ is
\[
\begin{split}
\Im \scr{M}(s) &= \frac{\ep^4}{2V}
\myint d^d x_1 \cdots d^d x_4 \,
e^{i(p_1 \cdot x_1 + \cdots + p_4 \cdot x_4)}
\\
&\qquad\qquad\qquad{}
\times \bra{0} \widebar{T}[\scr{O}(x_4) \scr{O}(x_3)]
T[\scr{O}(x_2) \scr{O}(x_1)] \ket{0},
\end{split}
\]
where the limits $p_3 \to -p_1$, $p_4 \to -p_2$ are understood.
We want to insert a complete set of states in the matrix element
between the $\widebar{T}[\cdots]$ and $T[\cdots]$ factors.
These states can be described using the
same operator-state correspondence used in Euclidean position space.
We give the main ideas here, and provide details in Appendix~\ref{sec:states}.

We begin with the radial quantization state defined by
inserting a primary scalar operator at the origin in Euclidean space:
\[
\ket{\Psi} = \Psi_\text{E}(0) \ket{0}
\]
These states are defined on a Hilbert space that lives on
a unit sphere in Euclidean space.
We then use a combination of conformal transformation and
Wick rotation to map this sphere into the $t = 0$ slice of Minkowski
space.
We view the conformal transformation as acting on the operators and not
the states (``Heisenberg picture''), so this tells us how to interpret the
same state $\ket\Psi$ in terms of Minkowski space correlation functions.%
\footnote{The state $\ket\Psi$ is not primary in Minkowski space because
the Minkowski special conformal generator $K_\mu$ is not the same as the Euclidean
special conformal generator.
Details are in Appendix~\ref{sec:states}.}
The result is that the state $\ket\Psi$ is given by inserting the operator
$\Psi$ at a finite imaginary time, for example
\[
\bra{\Psi} T[\scr{O}(x_2) \scr{O}(x_1)] \ket{0}
&= \frac{\la_{{\cal OO}\Psi}}{(x_{10}^2)^{\De_{\Psi}/2} (x_{20}^2)^{\De_{\Psi}/2}
(x_{12}^2 + i \ep)^{\De_{\scr{O}} - \De_\Psi/2}},
\]
where $\la_{{\cal OO}\Psi}$ is an OPE coefficient
and $x_0 = (-i/2, \pvec{0})$.
(The magnitude of the imaginary time is arbitrary since it can be changed by
a dilatation.
We choose $-i/2$ for later convenience.)
Note that the finite imaginary part of the time component of
$x_0$ corresponds to inserting the operator $\Psi$
to the left of $T[\scr{O}(x_2) \scr{O}(x_1)]$.

The next step is to Fourier transform the state $\ket\Psi$ to define
an eigenstate of the Minkowski momentum operator $P_\mu$.
This is accomplished by
\[
\ket{\Psi(k)} = \myint d^d x\, e^{-i \, k \cdot x} e^{i \, P \cdot x} \ket{\Psi}.
\]
It is easy to see that this is the unique state in the conformal subspace
of $\ket\Psi$ that satisfies
$P_\mu \ket{\Psi(k)} = k_\mu \ket{\Psi(k)}$.
This means that by Fourier transforming the state $\ket\Psi$ we automatically
include the full contribution from the conformal subspace containing $\ket\Psi$.
We provide a proof of this statement in the appendix based on the insertion of a complete set of states in an arbitrary two-point function.
In this sense, the conformal blocks in Minkowski momentum space are
simply given by (squares of) Fourier-transformed 3-point functions.

For a general tensor operator we define states
\[
\ket{\Psi^I(k)} = \ep^I_{M}(k)
\ket{\Psi^{M}(k)},
\]
where $M = \mu_1 \cdots \mu_\ell$ denotes the Lorentz indices
and $\ep^I_M$ are a complete set of symmetric traceless polarization tensors.
The completeness relation for states can then be written as
\[
\eql{completeness}
\mathds{1} = \sum_\Psi \sum_I
\myint \frac{d^d k}{(2\pi)^d} \,
\th(k^0) \th(-k^2) \,
\frac{\ket{\Psi^I(k)}\bra{\Psi^I(k)}}{e^{-k^0} \, \Pi_\Psi(k)},
\]
where the step functions $\theta$ restrict the integral over $k$ to
physical momenta (momenta in the spectrum of the operator $P_\mu$).
The function $\Pi_\Psi$ measures the norm of the momentum-space state:
\[
\eql{normstates}
\braket{\Psi^I(k')}{\Psi^J(k)} = (2\pi)^d \de^d(k - k') \,
\theta\left( k^0 \right) \theta\left( -k^2 \right)
\de^{IJ} e^{-k^0}  \Pi_\Psi(k),
\]
where
\[
\eql{Piintegral}
\Pi_\Psi(k) \de^{IJ} =
\ep^I_M(k) \ep^J_N(k)
\myint d^d x\, e^{-i \, k \cdot x}
\bra{0} \Psi^M(x)
\Psi^N(0) \ket{0}.
\]
Note that the norm of the states \Eq{normstates}
is not Lorentz invariant because we have
chosen a time slice $x^0 = 0$ to define the states.
The norm can be written in terms of the Lorentz invariant function
$\Pi_\Psi(k)$ by shifting the $x^0$ integral by $i$ in \Eq{Piintegral}.
Of course, we will find that all Lorentz non-invariance cancels when computing
Lorentz invariant quantities.

\subsection{Computation of the Imaginary Part of the Amplitude}
\scl{ImM}

We now discuss the calculation of the imaginary part of the amplitude, and subsequently of the coefficients $C_\Psi$ appearing in \Eq{csumrule}.

Using the completeness relation \eq{completeness} with a sum over all operators appearing in the $\scr{OO}$ OPE, the imaginary part of the amplitude is given by
\[
	\eql{ImM}
	2\Im\mathcal{M}(s)
	= \epsilon^4 \sum_\Psi \sum_I
	\left. \frac{|\la_{\scr{OO}\Psi}
	V_\Psi^I(p_1, p_2)|^2}{\Pi_\Psi(p_1 + p_2)}
	\right|_{p_1^2,\, p_2^2 \, = \, 0}
\]
where
\[
\la_{\scr{OO}\Psi} V_\Psi^I(p_1, p_2)
= \ep^I_M(p_1 + p_2)
&\myint d^d x_1 \, d^d x_2 \,
e^{i (p_1 \cdot x_1 + p_2 \cdot x_2)}
\bra{0} \Psi^M(0) T[ \scr{O}(x_1) \scr{O}(x_2)]\ket{0}.
\eql{Vintegral}
\]
The ordering in the 3-point function must be enforced by the correct
$i\ep$ prescription (see \Eq{VPsi} below). Then the coefficient $C_\Psi$ is given by
\[
C_\Psi = \frac{2}{\pi} \sum_I \left. \frac{|V_\Psi^I(p_1, p_2)|^2}{\Pi_\Psi(p_1 + p_2)}\right|_{p_1^2,\, p_2^2 \, = \, 0} ,
\]
which is manifestly non-negative.

These integrals can be carried out explicitly, and are finite in general.
For example, for a scalar operator $\Psi$ we have
\[
\eql{PiPsi_scalar}
\Pi_\Psi(p_1 + p_2) &= \frac{{2^{d - 2\Delta_\Psi + 1} {\pi ^{(d+2)/2}}}}
{\Ga \left( \De_\Psi  \right)
\Ga \left( \De_\Psi  - \frac{d-2}{2} \right)}
s^{\De_\Psi - d/2}
\]
and
\[
V_\Psi(p_1 + p_2) = & \myint d^d x_1 \, d^d x_2 \, e^{i(p_1 \cdot x_2 + p_2 \cdot x_2)}
\nn
&\qquad
\times
\frac{1}
{(x_{12}^2 + i\ep)^{\De_{\scr{O}} - \De_\Psi/2}
\bigl[ \vec{x}_1^{\gap 2} - (x_1^0 - i\ep)^2 \bigr]^{\De_\Psi / 2}
\bigl[ \vec{x}_2^{\gap 2} - (x_2^0 - i\ep)^2 \bigr]^{\De_\Psi / 2}}
\nonumber\\[4pt]
= & \, \frac{-i \, 2^{2d - 2 \De_\scr{O} - \De_\Psi + 1} {\pi ^{d + 1}}
\bigl[ {\Ga\left( {\De_{\cal O}} - \frac{d}{2} \right)} \bigr]^2
\Ga\left( \frac{d}{2} - \De_{\cal O} + \frac{\De_\Psi}{2} \right) {s^{{\De_{\cal O}} + {\De_\Psi }/2 - d}}}
{\bigl[ \Ga \left( \frac{\De_\Psi}{2} \right) \bigr]^2
\Ga\left( {\De_{\cal O}} - \frac{\De_\Psi}{2} \right)
\Ga\left( \De_{\cal O} + \frac{\De_\Psi}{2} - \frac{d}{2} \right)
\Ga\left( \De_{\cal O} + \frac{\De_\Psi}{2} - d + 1 \right)}.
\eql{VPsi}
\]
The dependence of the functions $V_\Psi^I$ and $\Pi_\Psi$ on $s$ cancels by scale invariance when $\Delta_\scr{O} = 3d/4$, so that the coefficient $C_\Psi$ of \Eq{csumrule} is a pure number that can be written, for a scalar intermediate operator,
\[
	\eql{CPsiscalar}
	C_\Psi = \frac{2^{3d - 4 \De_\scr{O} + 2} \pi^{3d/2}
	\Ga \left( \De_\Psi  \right)
	\Ga \left( \De_\Psi  - \frac{d-2}{2} \right)
	\bigl[ {\Ga\left( \frac{d}{4} \right)} \bigr]^4
	\bigl[ \Ga\left( \frac{\De_\Psi}{2} - \frac{d}{4} \right) \bigr]^2 }
	{\bigl[ \Ga \left( \frac{\De_\Psi}{2} \right) \bigr]^4
	\bigl[ \Ga\left( \frac{3d}{4} - \frac{\De_\Psi}{2} \right)
	\Ga\left( \frac{\De_\Psi}{2} + \frac{d}{4} \right)
	\Ga\left( \frac{\De_\Psi}{2} - \frac{d}{4} + 1 \right) \bigr]^2},
\]

\begin{figure}
	\centering
	\includegraphics[scale=0.8]{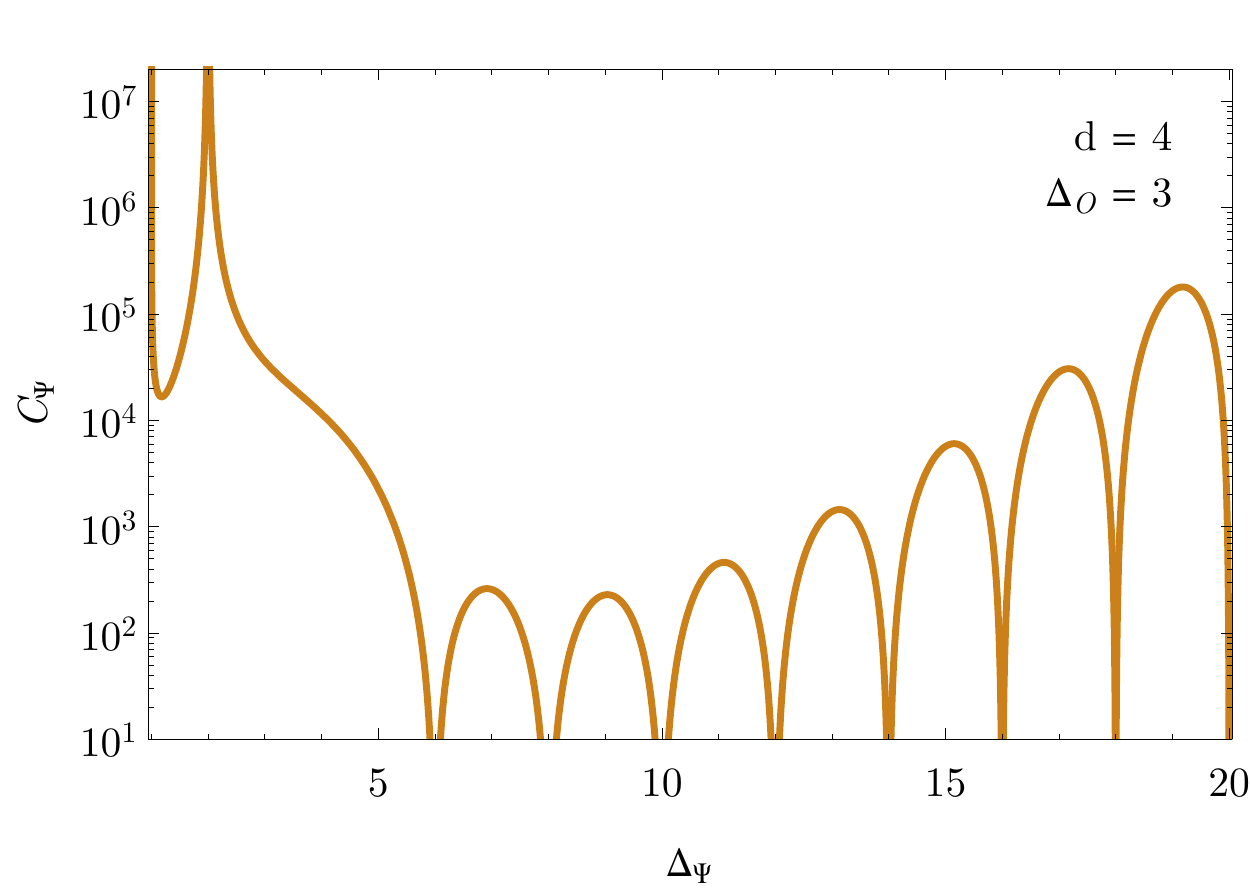}
	\caption{The coefficient $C_\Psi$ of \Eq{CPsiscalar} for a scalar intermediate operator $\Psi$ in $d = 4$. The zeroes of $C_\Psi$ when $\Delta_\Psi = 2\De_{\scr{O}} + 2n$ are visible, as are the UV divergences at $\Delta_\Psi = d/2$ and $(d-2)/2$. At large $\Delta_\Psi$, the coefficient $C_\Psi$ increases exponentially.}
	\label{fig:momentum_space_coeffs}
\end{figure}

The coefficients $C_\Psi$ are manifestly non-negative,
and are non-singular except at $\Delta_\Psi = d/2$.
They vanish when $\Psi$ has the dimension of a double trace operator, namely
\[
\De_\Psi = 2\De_{\scr{O}} + 2n,
\qquad
n = 0, 1, 2, \ldots
\]
This must in fact be the case, as our
sum rule must give zero for a generalized free field theory, where the 4-point
function has no connected part.
Such a theory however has a nontrivial OPE with an infinite number of
double trace operators, and positivity means that the only way to satisfy the
sum rule in this case is to have $C_\Psi = 0$ for any scalar operator
with double trace dimension.
One can also see in \Eqs{VPsi} and \eq{CPsiscalar}
the presence of the UV divergence discussed in \sec{UVdivergences}, at $\Delta_\Psi = 2 \Delta_\scr{O} - d$, saturating the bound of \Eq{UVdivergencebound}.
A plot of $C_\Psi$ as a function of $\De_\Psi$ in $d = 4$ is given in \Fig{momentum_space_coeffs}.

We have not computed the coefficients $C_\Psi$ for a general spin-$\ell$ operator,
although this could be done with additional labor.
We have calculated them for the special case where the spin-$\ell$ operators
saturate the unitarity constraints, and are therefore higher-spin
conserved currents.
This is the case that appears in the free-field example, to which we turn next.

\subsection{A Worked Example}

We can apply the momentum space sum rule to the free scalar field theory in
$d = 8$ with $\scr{O} = \phi^2 /\sqrt{2}$.
This theory has a scale anomaly of the form \Eq{thescaleanomexample}, and one can check that
there are no UV or IR divergences that invalidate the sum rule.

The $\phi^2 \phi^2$ OPE contains operators of the form $\sim \d^{2n} \phi^4$,
but these have double-trace dimension and therefore do not contribute to the
sum rule.
(The agrees with the fact that in terms of Feynman diagrams there are no
connected diagrams for $AA \to \phi\phi\phi\phi$.)
Because the theory is invariant under $\phi \mapsto -\phi$, the only other
operators we need to consider have the form $\d^{2n} \phi^2$.
Moreover, only operators with spin $\ell = 2n$ are primaries, and since they saturate the unitarity bound, they must be conserved higher-spin currents.
We provide the details of the computation of the two- and three-point function in Appendix~\ref{sec:freescalar}. The result can be written as a sum over $n$,
\[
	c_{\scr{O}} = \frac{\pi^{12}}{4}
	\sum_{n=0}^\infty \frac{4n+5}{(n+1) (n+2) (2n+1) (2n+3)}
	= \frac{\pi^{12}}{4}.
\]
This sum reproduces the position-space result of \Eq{freescalaranomaly}
as well as a simple computation using Feynman diagram in \Eq{cphi2:2}.
The convergence does not seem to be as fast as in the position space case (the convergence is only linear, not exponential), but the positive nature of the sum makes it much more valuable for future use.

\section{Conclusions}
In this paper we investigated the scale anomaly of the 4-point functions
for scalar operators in
a general CFT, and showed, with some caveats, that the anomaly coefficient
can be expressed in the form of an OPE, namely
\[
c_{\scr{O}} = \sum_\Psi \la_{\scr{OO}\Psi}^2 C_\Psi,
\]
where the sum runs over primary operators $\Psi$ that appear in the
$\scr{OO}$ OPE.
The most interesting version of this sum rule came from thinking of the
$\avg{\scr{OOOO}}$ correlation function as a scattering amplitude in
Minkowski momentum space.
In this version, the identity operator does not appear, and the coefficients
$C_\Psi$ are non-negative.
We developed a calculus for Minkowski correlation functions that allowed
us to write a convergent OPE in momentum space.
The main caveat to the momentum-space anomaly sum rule is that we do not have
a rigorous understanding of the absence of IR divergences in the amplitude.

We also wrote a version of this sum rule that follows from the correlation
function in Euclidean space.
In this version, the identity operator does appear, and the coefficients
$C_\Psi$ do not have a definite sign.

There are a number of extensions and directions for further investigations:
\begin{itemize}
\item
Understand better the IR convergence of the
momentum space sum rule.
\item
Extend the results to operators with spin, in particular conserved currents and the energy-momentum tensor,
and  obtain a sum rule for the anomaly coefficient $c_T$ in 4-dimensional CFTs.
\item
Extend this approach to the $a$ anomaly coefficient of a 4-dimensional
conformal field theory.
This anomaly does not lead to an anomaly for scale transformations, but does give
rise to an anomaly for special conformal transformations.
It is not clear that one should expect positivity for such a sum rule,
if one can be found.
\item
Investigate other possible uses of the calculus we have
developed for the OPE for Wightman ordered products in Minkowski space.
\end{itemize}
We hope to address these questions in future work.

\subsection*{Acknowledgments}
We have benefited from discussions and encouragement from many people, including
S. Dubovsky,
J. Kaplan,
Z. Khandker,
R. Rattazzi,
S. Rychkov,
R. Sundrum,
and M. Walters.
M.G.~is supported by the
Swiss National Science Foundation under grant number P300P2154559
and by NCCR SwissMAP.
X.L.~and M.A.L.~are supported by
the Department of Energy under grant DE-FG02-91ER406746.


\appendix
\titleformat{\section}{\centering\normalsize\bfseries}{Appendix~\thesection:~}{0pt}{}

\section{Radial Quantization States in Momentum Space}
\scl{states}

In this appendix we describe how to use the state-operator correspondence
to write a complete set of states in Minkowski space.
We start with radial quantization in Euclidean space, and use 
a combination of conformal transformations and Wick rotation
to relate this to states defined on a time slice $x^0 = 0$ in Minkowski space.
This allows us to write matrix elements of Minkowski space operators with
radial quantization states. 
We also give an independent argument that these states are complete by using
them to reproduce the 2-point functions of the theory.

\subsection{Conformal Mapping}
Let us recall some basics of radial quantization in CFT.
In any quantization, we pick a time surface and define ``in'' (``out'') states
by inserting operators to the past (future) of the surface into the path integral.
Correlation functions can then be viewed as the overlap of in and out states:
\[
\avg{\scr{O}_1 \cdots \scr{O}_n}
= {}_\text{out} \braket{\scr{O}_1 \cdots \scr{O}_k}
{\scr{O}_{k+1} \cdots \scr{O}_n}_\text{in}.
\]
We begin with radial quantization in Euclidean space.
We denote the standard Euclidean coordinates by $x^\mu$ ($\mu = 1, \ldots, d$) and define
radial quantization states on the unit sphere $x^2 = 1$ by inserting
operators at the origin and integrating over the region $x^2 < 1$:
\[
\ket\Psi = \ket{\Psi(x = 0)}\!\!{}_{\substack{\text{in} \\ x^2 < 1}}. 
\]
For example, 2- and 3-point functions in radial quantization are given by 
\[
{}_{\substack{\text{out} \\ x^2 > 1}} \! \braket{\Psi(x_1)}{\Psi} &= \avg{\Psi(x_1) \Psi(x = 0)}
= \frac{1}{|x_1|^{2\De_\Psi}},
\\
{}_{\substack{\text{out} \\ x^2 > 1}} \! \braket{\scr{O}(x_1) \scr{O}(x_2)}{\Psi}
&= \avg{\scr{O}(x_1) \scr{O}(x_2) \Psi(x = 0)}
= \frac{\la_{\scr{OO}\Psi}}{|x_{12}|^{2\De_{\scr{O}} - \De_\Psi} |x_1|^{\De_\Psi} |x_2|^{\De_\Psi}}.
\]
In radial quantization,
Hermitian conjugation is given by inversion.
This makes sense because inversion maps the path integral over fields inside the sphere to
the path integral outside the sphere.
For example, we have
\[
\bra\Psi = \lim_{x \to 0} x^{-2\De_\Psi}
{}_{\substack{\text{out} \\ x^2 > 1}} \! \bra{\Psi(x^{-1})},
\]
where $(x^{-1})^\mu = x^\mu / x^2$.
Note that this implies
\[
\braket\Psi\Psi &= \lim_{x \to 0} x^{-2\De_\Psi} \frac{1}{|x^{-1} - 0|^{2\De_\Psi}} = 1.
\]

Any conformal transformation $x^\mu \mapsto x'^\mu$ maps radial quantization operators
and states to operators and states in a new quantization.
In radial quantization, in-states are defined by integrating over the region $x^2 < 1$,
while in the new quantization, in-states are defined by integrating over the image
of this region.
%
%
%
%
%
%
%
%
We apply these ideas to a particular conformal map that maps the unit sphere
$x^2 = 1$ into the plane $x'^d = 0$.
This gives a quantization where the states live on the plane $x'^d = 0$.
The transformation is a combination of a translation and a special conformal transformation,
and is given explicitly by
\[
\pvec{x}' = R \frac{2\vec{x}}{1 + 2 x^d + x^2},
\qquad
x'^d = R \frac{x^2 - 1}{1 + 2 x^d + x^2}.
\]
where $R$ is an arbitrary length scale.
(We think of the unprimed variables $x^\mu$ as dimensionless.)
Note that the origin is mapped to $x'_0 = (\vec{0}, -R)$,
and the point at infinity is mapped to $x'_\infty = (\vec{0}, R)$.
Conformal symmetry implies
\[
{}_{\substack{\text{out} \\ x^2 > 1}} \! \braket{\scr{O}(x_1) \scr{O}(x_2)}{\Psi}
&= \avg{\scr{O}(x_1) \scr{O}(x_2)\Psi(0)}
\nn
&= \avg{\scr{O}'(x'_1) \scr{O}'(x'_2)\Psi'(x'_0)}
\nn
\eql{quantizexpl}
&= {}_{\substack{\text{out} \\ x'^d > 0}} \! \braket{\scr{O}'(x'_1) \scr{O}'(x'_2)}{\Psi'},
\]
where 
\[
\ket{\Psi'} = 
\ket{\Psi'(x'_0)}\!\!{}_{\substack{\text{in}\\x'^d < 0}},
\qquad
\scr{O}'(x') = \Om^{-\De_{\scr{O}}}(x) \scr{O}(x).
\]
Because the conformal transformation is unitary, we have $\ket{\Psi'} = U \ket{\Psi}$,
that is, this is a unitary transformation from one basis to another.
In the primed quantization, we therefore have
\[
{}_\text{out}\braket{\scr{O}'(x'_1) \scr{O}'(x'_2)}{\Psi'}
&= \frac{\la_{\scr{OO}\Psi}}
{x'^{2\De_{\scr{O}} - \De_\Psi}_{12} x'^{\De_\Psi}_{10} x'^{\De_\Psi}_{20}}.
\]
In other words, in this quantization, the states $\ket{\Psi'}$ are defined by
inserting an operator at the point $x'_0$.
%

It is easily seen that inversion in the unprimed variables corresponds to reflection
about $x'^d = 0$:
\[
x \mapsto x^{-1} 
\quad\Leftrightarrow\quad
(\pvec{x}', x'^d) \mapsto (\pvec{x}', -x'^d).
\]
This means that Hermitian conjugation is simply given by
\[
\left[ \ket{\Psi'(\pvec{x}', x'^d)}_\text{in} \right]^\dagger
= {}_\text{out}\bra{\Psi'(\pvec{x}', -x'^d)}.
\]
In particular, we have
\[
\braket{\Psi'}{\Psi'} = \frac{1}{(2R)^{2\De_\Psi}}.
\]
We see that it is convenient to choose $R = \frac 12$ to avoid factors of 2 in the
normalization of states.

\subsection{Wick Rotation}
Note that we have not specified a Hamiltonian for the primed quantization above.
In radial quantization, the Hamiltonian is the dilatation operator.
If we use the image of the dilatation operator under this conformal mapping
as the Hamiltonian in the primed quantization, we obtain the so-called ''N-S quantization'' \cite{Rychkov:2016iqz}.
But we can use any Hamiltonian we like as long as the surface $x'^d = 0$ is
a surface of constant ``time.''
For our purposes it is most useful to use $x'^d$ as the time variable.

We then define the Wick rotation by
\[
x'^0 = -i x'^d,
\]
and let $\mu = 0, 1, \ldots, d-1$.
Now our states $\ket{\Psi'}$ live on the $x'^0 = 0$ time slice of Minkowski space. 

The Minkowski space path integral then defines the matrix element
\[
&{}_\text{out}\braket{\scr{O}(x'_1) \scr{O}(x'_2)}{\Psi}
= \bra{0} T[\scr{O}(x'_1) \scr{O}(x'_2)] \ket{\Psi}
\\
& \qquad
= \frac{\la_{\scr{OO}\Psi}}{\bigl[\pvec{x}'^2_{\! 12} - (x_{12}'^0)^2 + i\ep\bigr]^{\De_\Psi - \De_{\scr{O}}/2}
\bigl[\pvec{x}'^2_{\! 1} - (x_1'^0 - iR)^2\bigr]^{\De_\Psi/2}
\bigl[\pvec{x}'^2_{\! 2} - (x_2'^0 - iR)^2\bigr]^{\De_\Psi/2}}.
\nonumber
\]
Note that the path integral now gives the time ordered product of operators in the out
state, so the 3-point function has the standard Feynman $i\ep$ prescription
$x_{12}'^2 \to \pvec{x}'^2_{\! 12} - (x_{12}'^0)^2 + i\ep$.
The terms involving $x_0 = (iR, \pvec{0})$ involve imaginary time and do not
need any additional $i\ep$ prescription.
The fact that $\Psi$ is at a finite positive imaginary time ensures that it is inserted
to the right of all the other operators.
Note also that after the Wick rotation, Hermitian conjugation becomes equivalent to ordinary complex conjugation. The conjugate state $\bra{\Psi}$ correspond therefore to a local operator insertion at $x_0^* = (-iR, \pvec{0})$, to the left of all other operators.

\subsection{Alternative Derivation}

\begin{figure}
	\centering
	\includegraphics[width=3cm]{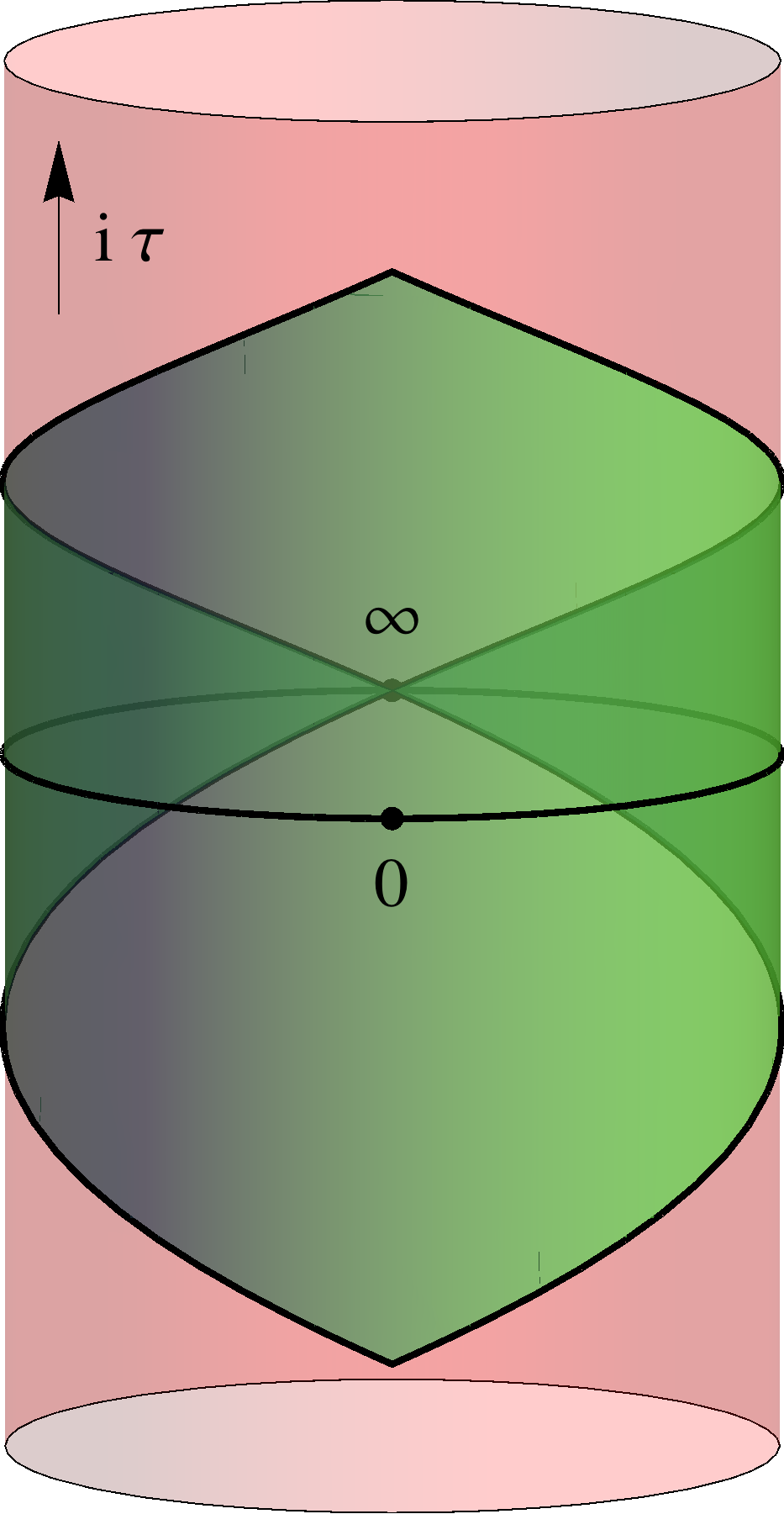}
	\caption{The Poincar\'e patch in the cylinder coordinates, with $i \tau = i \log|x|$ running vertically and $\vec{x}/|x|$ represented by an angle in the horizontal plane. The points labeled 0 and $\infty$ correspond to the origin and spatial infinity of Minkowski space, and both of them lie on the quantization surface $\tau = 0$.}
	\label{fig:Poincarepatch}
\end{figure}

There are other ways to obtain the results above through a combination of conformal
transformations and Wick rotations.
One way that is perhaps more familiar is to first map radial quantization to the
Euclidean cylinder by $r = e^{\tau}$,
then perform the Wick rotation of the cylinder time $t = i\tau$.
This gives a theory on the Minkowski cylinder,
and states on the radial quantization sphere $r = 1$
are mapped to the sphere at $t = 0$ in this quantization.
To obtain flat Minkowski space, we can then restrict the theory to the Poincar\'e
patch shown in \Fig{Poincarepatch} using a special conformal transformation.
This approach yields precisely the same final expressions as above.
Conceptually it is a little less clear to us since the origin
$r = 0$ is mapped to $t \to i\infty$, but then comes back to a finite (imaginary) point only after a final change of variables.

Yet another approach is to define the state $\ket{\Psi}$ in
Minkowski space directly by inserting $\Psi$ at $x_0' = (-iR, \pvec{0})$. 
It can then be seen that $\ket{\Psi}$ is annihilated by the operators
\[
	K_i = -\dfrac{R \, P'_i - R^{-1} K'_i}{2} - i M'_{0i},
	\hspace{1.5cm}
	K_d = i \dfrac{R \, P'_0 + R^{-1} K'_0}{2} + D',
\]
and is an eigenstate of
\[
	\eql{Hamiltonian}
	D = i \dfrac{R\, P'_0 - R^{-1} K'_0}{2},
\]
with eigenvalue $\Delta_\Psi$, where the operators $P'_\mu$, $K'_\mu$, $M'_{\mu\nu}$ and $D'$ represent the generators of the $\SO{d,2}$ conformal group in the primed coordinates. $K_\mu$ and $D$ are part of a representation of the Euclidean conformal group $\SO{d+1,1}$ acting on the unprimed coordinates of the previous section.
They can be used to construct the full spectrum of states in the theory. Descendants states in particular are obtained by acting on $\ket{\Psi}$ with the operators $P_\mu$ conjugate to $K_\mu$, given by
\[
	P_i = \dfrac{R \,P'_i - R^{-1} K'_i}{2} - i M'_{0i},
	\hspace{1.5cm}	
	P_d = i \dfrac{R \, P'_0 + R^{-1} K'_0}{2} - D'.
	\eql{raisingoperators}
\]
This quantization procedure is the one described by L\"uscher and Mack~\cite{Luscher:1974ez}. Its Hamiltonian is given in terms of Minkowski coordinates $x'_\mu$ by \Eq{Hamiltonian}.

\subsection{Momentum space states}

Where our quantization method differs from L\"uscher and Mack's is in the construction of descendant states. Acting on the primary states $\ket{\Psi}$ with the raising operators $P_\mu$ of \Eq{raisingoperators} generates all the states in the theory, but these states are not convenient to use. In particular they are not momentum eigenstates.
If instead one tries to generate descendant states by acting on a primary state with the momentum operator $P'_\mu$, then the set of states will still be complete, since the relation between $P'_\mu$ and $P_\mu$ is linear, but the states obtained in this way are not orthogonal to each other, as $\ket{\Psi}$ is not an eigenstate of the dilatation operator $D'$.

An elegant solution to this problem is obtained taking the Fourier transform
\[
	\eql{states:momentumspace}
	\ket{\Psi(k)}
	= & \myint d^dx' \, e^{-i \, k \cdot x'}
	e^{i \, P' \cdot x'}
	\ket{\Psi}
	\nonumber \\
	= & \myint d^dx' \, e^{-i \, k \cdot x'}
	\Psi( x'^0 - i R, x'^i ) \ket{0}.
\]
The states obtained in this way are remarkable in many ways: they have non-negative norm, are orthogonal to each other (see \Eq{normstates} and \Eq{Piintegral}), and are eigenstates of the translation operator $P'_\mu$ with eigenvalue $k_\mu$. Moreover, their overlap with in or out states is finite due to the presence of the imaginary time component. These properties allow to write a closure relation in the form of \Eq{completeness}, which is correct provided that the set of states $\ket{\Psi(k)}$ with $k^\mu$ in the future light-cone is complete. This last point is not trivial, but we show its proof in the next section. Note that we have only mentioned scalar operators so far. The generalization to operators with spin is trivial, and the completeness equation \eq{completeness} readily incorporates them.

\subsection{Two-point functions and completeness of the set of states}

Let us consider the Wightman two-point function for a scalar operator $\scr{O}$ in Minkowski position space,
\[
	\eql{2ptfunction}
	\bra{0} \scr{O}(x) \scr{O}(0) \ket{0}
	= \frac{1}{\left[ (x^i)^2 - (x^0 - i \epsilon)^2 \right]
	^{\Delta_\scr{O}}}.
\]
In the Euclidean version of this two-point function, one can compute the \rhs\ by inserting between $\scr{O}(x)$ and $\scr{O}(0)$ radial quantization states defined on a sphere of radius $0 < R < |x|$. It is a known result that all the states in the conformal class of $\scr{O}$ must be taken into account to recover the full two-point function at finite separation $x$. This means that if we can reproduce the same result in Minkowski space inserting our set of states \Eq{completeness} on the \lhs\ of \Eq{2ptfunction}, we will have proven that our basis is complete, at least in the conformal class of $\scr{O}$. The generalization to fields with spin follows straightforwardly.

Plugging \Eq{completeness} in the above two-point function and using translation invariance to factorize the $x$ dependence, one obtains the Fourier transform of a positive quantity,
\[
	\bra{0} \scr{O}(x) \scr{O}(0) \ket{0}
	= & \myint \frac{d^d k}{(2 \pi)^d} \,
	e^{-i \, k \cdot x} \,
	\theta(k^0) \theta(-k^2) \,
	\frac{\bigl| \bra{0} \scr{O}(0) \ket{\scr{O}(k)} \bigr|^2}
	{e^{-k^0} \Pi_{\scr{O}}(k)}
	\nonumber \\
	= & \myint \frac{d^d k}{(2 \pi)^d} \,
	e^{-i \, k \cdot x} \,
	\theta(k^0) \theta(-k^2) \,
	\Pi_{\scr{O}}(k),
\]
where in the second equality we have used the definition of $\Pi_\scr{O}(k)$ (see \Eq{PiPsi_scalar}) to rewrite the integrand in a manifestly Lorentz-invariant way. The last line turns out to be the Fourier transform of the usual Wightman propagator in momentum space, reproducing precisely the \rhs\ of \Eq{2ptfunction}, hence concluding the proof.



\section{Computing the Flux Integral in Position Space}
\scl{fluxintegral}

We detail in this appendix the computation of the flux integral in \Eq{cfluxint}. As mentioned before, the idea is to make use of conformal symmetry to simplify the integral. By definition, the integral is over the four points $x_1$ to $x_4$, but  the four-point function only depends non-trivially on two real parameters: using conformal invariance, three of the four points can be brought to the points $(0, \ldots, 0)$, $(1, 0, \ldots, 0)$ and $\infty$, and the position of the last point is then only parametrized by its distance to the former two points. The integral should therefore be reducible to a two-dimensional space. We detail below this procedure, starting with the definition \Eq{cfluxint} and leading to the result \Eq{c_position_space}.

The first step is to use translation invariance of the integrand to set $x_4 = 0$, and for later convenience to write the surface integral as a volume integral with a delta function,
\[
	c_\scr{O} = & \myint d^dx_1 d^dx_2 d^dx_3 \,
	\delta\left(\sqrt{x_1^2 + x_2^2 + x_3^2} - 1 \right)
	\nonumber \\
	& \qquad \times
	\frac{G_\scr{O}(u,v)}
	{\left[ x_1^2 x_2^2 x_3^2 x_{12}^2 x_{13}^2
	x_{23}^2  \right]^{d/4}}
\]
where $G_\scr{O}(u,v)$ is related to the sum over conformal blocks of \Eq{confblockexpansion}, minus its disconnected part,
\[
	G_\scr{O}(u,v) = \left( \frac{v}{u^2} \right)^{d/4}
	\sum_\Psi \lambda_{\scr{O}\scr{O}\Psi}^2
	g_\Psi (u, v)
	- \left( \frac{v}{u^2} \right)^{d/4}
	- \left( u v \right)^{d/4}
	- \left( \frac{u}{v^2} \right)^{d/4} .
\]
The second step consists in moving $x_1$ to infinity using the Faddeev-Popov method: one starts with inserting the identity in the integral in the form of $\myint d^db \, \delta^d \left( b^\mu - x_1^\mu/x_1^2 \right)$,
and then one performs the change of variable
\[
	x_i^\mu \to \frac{x_i^\mu + b^\mu x_i^2}
	{1 + 2 b \cdot x_i + b^2 x_i^2}.
\]
Upon integration over $x_1$, the previously inserted delta function selects the point $x_1 \to \infty$, and we obtain
\[
	c_\scr{O} = & \myint d^db \, d^dx_2 \, d^dx_3 \,
	\frac{G_\scr{O}(u,v)}
	{\left[ x_2^2 x_3^2 x_{23}^2  \right]^{d/4}}
	\left[ b^2 \left( 1 + 2b \cdot x_2 + b^2 x_2^2 \right)
	\left( 1 + 2b \cdot x_3 + b^2 x_3^2 \right) \right]^{-d/4}
	\nonumber \\
	& \qquad\qquad \times
	\delta\left(\sqrt{\frac{1}{b^2}
	+ \frac{x_2^2}{1 + 2b \cdot x_2 + b^2 x_2^2}
	+ \frac{x_3^2}{1 + 2b \cdot x_3 + b^2 x_3^2}} - 1 \right).
\]
The next step is to use a scale transformation to set the norm of $x_2$ to unity, using again the Faddeev-Popov method. Inserting $\int_0^\infty d\lambda \, \delta(\lambda - |x_2|)$ in the integral and performing the change of variables $x_i \to \lambda x_i$, $b \to \lambda^{-1} b$, we get after successive integration over $x_2$ and $\lambda$,
\[
	c_\scr{O} = \frac{2 \pi^{d/2}}{\Gamma\left( \frac{d}{2} \right)}
	\myint d^db d^dx_3 \,
	\frac{G_\scr{O}(u,v)}
	{(u v)^{d/4}}
	\left[ b^2 \left( 1 + 2b \cdot n + b^2 \right)
	\left( 1 + 2b \cdot x_3 + b^2 x_3^2 \right) \right]^{-d/4},
\]
where $n = (1,0,\ldots, 0)$ is an arbitrary unit vector.
The only remaining conformal invariance of this integral is in the form of $\SO{d-1}$ rotations in the directions orthogonal to $n$. Denoting $\vec{z} = (z_1, z_2, 0, \ldots 0)$ and $z = z_1 + i z_2$, the integral over $x_3$ can be put in the form of \Eq{c_position_space},
\[
	c_\scr{O} = \int\limits_{\Im(z) > 0}
	\frac{d z d \zbar}{(\Im z)^2}
	\scr{K}_d\left( z, \zbar \right)
	G_\scr{O}(u,v),
\]
where the conformal cross-ratios are now $u = |z|^2$, $v = |1-z|^2$, and the integration kernel is given by
\[
	\scr{K}_d\left( z, \zbar \right)
	= & \, \frac{4 \, \pi^{(2d - 1)/2}}
	{\Gamma\left( \frac{d-1}{2} \right)
	\Gamma\left( \frac{d}{2} \right)}
	\frac{|\Im z|^d}{|z|^{d/2} |1-z|^{d/2}}
	\nonumber \\
	& \qquad \times
	\myint d^db \,
	\frac{1}{\left[ b^2 \left( |z|^2 + 2|z| \, b \cdot n + b^2 \right)
	\left( 1 + 2 b \cdot \vec{z}/|z| + b^2 \right) \right]^{d/4}}.
\]
Using Feynman parameters to perform the integration over $b$, one obtains finally \Eq{int_kernel}. An equivalent formulation in term of the $_2F_1$ hypergeometric function is
\[
	\eql{int_kernel_2}
	\scr{K}_d \left( z, \zbar \right)
	= & \, \frac{4 \, \pi^{(3d-1)/2}}
	{\Gamma\left(\frac{d-1}{2} \right)
	\Gamma\left(\frac{d}{2} \right)^2}
	\frac{|\Im z|^d}{|z|^{d/2} |1-z|^{d/2}}
	\nonumber \\
	& \qquad
	\times \int_0^1 \frac{d s}{s(1-s)} \,
	_2F_1 \left( \frac{d}{4}, \frac{d}{4}; \frac{d}{2};
	- \frac{(z - s)(\zbar - s)}
	{s (1-s) } \right).
\]
This function has the following properties:
\begin{itemize}

\item
$\scr{K}_d$ is real and positive over the full complex $z$ plane, symmetric with respect to complex conjugation $z \to \zbar$, and has two absolute maxima at the completely crossing-symmetric points $z = \frac{1 \pm i\sqrt{3}}{2}$ (corresponding to $u = v = 1$).

\item
$\scr{K}_d$ vanishes on the real axis ($z = \zbar$), and so does $\scr{K}_d/(\Im z)^2$. This property actually regulates the branch cut of the conformal block $g_\scr{O}(u,v)$ at $z > 1$.

\item
$\scr{K}_d$ is continuous everywhere but not differentiable at the points $z = 0$ and $z = 1$, where it has the limiting behavior given in \Eq{Katzero}.

\end{itemize}
When $d$ is an integer multiple of 4, the hypergeometric function in \Eq{int_kernel_2} reduces to a logarithm times a rational function, and the integral can be performed explicitly. In $d = 4$, for instance, we find the simple result
\[
	\eql{int_kernel_4d}
	\scr{K}_4 \left( z, \zbar \right)
	= 16 \pi^5
	\frac{(\Im z)^3}{|z|^2 |1-z|^2}
	\big[ \Im\left( \Li_2(z) \right)
	+ \arg(1-z) \log|z| \big],
\]
where the term in square bracket is the Bloch-Wigner dilogarithm~\cite{Zagier:1988}, a function that is known to appear in various places in physics.%
\footnote{For instance in supersymmetric multi-loop amplitudes~\cite{Eden:2000mv} or in AdS/CFT \cite{Arutyunov:2002fh}. The Bloch-Wigner dilogarithm has also a geometric interpretation as the volume of an ideal tetrahedron in hyperbolic space.
We thank Jo\~ao Penedones for pointing this out to us.}
In $d = 8$, the kernel is also proportional to the Bloch-Wigner dilogarithm,
\[
	\eql{int_kernel_8d}
	\scr{K}_8 \left( z, \zbar \right)
	= & \frac{4 \, \pi^{11}}{45}
	\frac{(\Im z)^4}{u^2 v^2}
	\nonumber \\
	& \times \bigg[ \left( 6 u v
	- 2 (\Im z)^2 (1 + u + v) \right)
	\frac{\Im\left( \Li_2(z) \right)
	+ \arg(1-z) \log|z|}{\Im z}
	\nonumber \\
	& \quad
	+ \left( 3 (1 - u)^2 - 3 v (1 + u)
	+ 8 (\Im z)^2 \right) \log|1-z|
	\nonumber \\
	& \quad
	+ \left( 3 u (1 - u + v)
	- 4 (\Im z)^2 \right) \log|z|
	-2 (\Im z)^2
	\bigg].
\]
A closed form result for $\scr{K}_d$ in arbitrary dimension is not known.


\section{The Free Scalar Theory in 8 Dimensions}
\scl{freescalar}

In this appendix, we provide more details about the free scalar theory in 8 dimensions, which is the simplest example of a theory with a scale anomaly in a scalar four-point function. We review in particular some known results about the free scalar theory in arbitrary $d$ and derive some identities that are being used in the main body of our work.

\subsection{The $\phi^2\phi^2$ OPE in Arbitrary $d$}

There are two types of operators that enter the $\phi^2\phi^2$ OPE, schematically of the form $\partial^{n} \phi^2$ and $\partial^{n} \phi^4$, where the derivative can actually act on any of the fields.
In general, such operators can have spin $\ell = n, n-2, \ldots$, where $n - \ell$ of the partial derivatives have contracted indices. For the operators of type $\partial^{n} \phi^2$, the situation is however simpler, as they cannot be primaries if $\ell \neq n$: they must either vanish by the equation of motion $\square \phi = 0$, if the two contracted derivatives act on the same $\phi$, or must be descendants. Therefore, all primary operators of type $\partial^n \phi^2$ have spin $\ell = n$, that is they are traceless symmetric tensors. Such operators saturate the unitarity bound and hence are conserved currents (except in the case $n = 0$). Using this conservation rule together with tracelessness and symmetry, one can show that there is a unique primary operator for each integer $n$ (see for instance \cite{Mikhailov:2002bp} or appendix F of~\cite{Penedones:2010ue}), which can be written as
\[
	\eql{phi2operators}
	\Psi_n^{\mu_1 \ldots \mu_n} = & \frac{1}{2 \, n!}
	\sum_{\nu_i \in \sigma\left( \mu_1, \ldots, \mu_n \right)}
	\sum_{s=0}^{\lfloor n/2 \rfloor}
	\sum_{k=0}^{n - 2s} c^{(n)}_{s,k} \,
	\eta^{\nu_1 \nu_2} \cdots
	\eta^{\nu_{2s-1} \nu_{2s}}
	\\
	& \hspace{2cm}
	\times \left( \partial^{\nu_{2s+1}} \cdots
	\partial^{\nu_{2s+k}}
	\partial^{\rho_1} \cdots \partial^{\rho_s} \phi \right)
	\left( \partial^{\nu_{2s+k+1}} \cdots
	\partial^{\nu_n}
	\partial_{\rho_1} \cdots \partial_{\rho_s} \phi \right).
	\nonumber
\]
The $c^{(n)}_{s,k}$ coefficients are fixed by the requirements of tracelessness and conservation to be of the form
\[
	\eql{doubletracecoeffs}
	c^{(n)}_{s,k} =
	\frac{(-1)^k \, n! \,
	\Ga\left( \frac{d-2}{2} \right)
	\Ga\left( \frac{d-2}{2} + n - s \right)}
	{2^s \, k! \, s! \, (n - 2s - k)! \,
	\Ga\left( \frac{d-2}{2} + s + k \right)
	\Ga\left( \frac{d-2}{2} + n - s - k \right)} \,
	c^{(n)}_{0,0}.
\]
This means in particular that there are no $\Psi_n$ operators with odd $n$, as the sum \Eq{phi2operators} vanishes in that case (which we know also from parity considerations). The coefficients $c^{(n)}_{0,0}$ are moreover fixed by the normalization condition
\[
	\big\langle \Psi_n^{\mu_1 \ldots \mu_n}(x_1)
	\Psi_n^{\nu_1 \ldots \nu_n}(x_2) \big\rangle
	= \frac{\scr{I}^{\mu_1 \nu_1}(x_{12}) \cdots
	\scr{I}^{\mu_n \nu_n}(x_{12})
	+ \text{permutations}
	- \text{traces}}
	{|x_{12}|^{2d + 2n -4}}
\]
where
\[
	\scr{I}^{\mu \nu}(x) = \eta^{\mu\nu}
	- \frac{2 \,  x^\mu x^\nu}{x^2},
\]
to
\[
	c^{(n)}_{0,0} = \left[ \frac{n!}{2^{n-1}}
	\frac{\Ga(d + n - 3)}{\Ga(d + 2n -3)}
	\right]^{1/2}.
\]
The first operator in this series is $\Psi_2 = \phi^2/\sqrt{2}$ itself, and the second coincides with the stress-energy tensor for the free scalar theory.

The operators $\Psi_{2n}$ are also the double-trace operators appearing in the OPE for the free scalar field $\phi$. If we denote by $\lambda_{\phi\phi\Psi_{2n}}$ the OPE coefficient of the three-point function
\[
	\eql{phiphiOn}
	\left\langle \Psi_{2n}^{\mu_1 \ldots \mu_{2n}}(x_1)
	\phi(x_2) \phi(x_3) \right\rangle
	= \frac{\lambda_{\phi\phi\Psi_{2n}}}
	{|x_{12}|^{d-2}
	|x_{13}|^{d-2}}
	\left( Z^{\mu_1} Z^{\mu_2} \cdots Z^{\mu_{2n}}
	- \textrm{traces} \right),
\]
where
\[
	Z^\mu = \frac{x_{12}^\mu}{x_{12}^2}
	- \frac{x_{13}^\mu}{x_{13}^2},
\]
then with the above definition of the operator $\Psi_{2n}$ the OPE coefficients take the form~\cite{Fitzpatrick:2012yx}
\[
	\eql{OPEcoefficients}
	\lambda_{\phi\phi\Psi_{2n}}
	= \left[ \frac{2^{2n+1}}{(2n)!}
	\frac
	{\Ga\left( \frac{d-2}{2} + 2n \right)^2
	\Ga(d + 2n - 3)}
	{\Ga\left( \frac{d-2}{2} \right)^2
	\Ga(d + 4n -3)} \right]^{1/2}.
\]
The three-point function of two operators $\phi^2$ (or rather $\Psi_2$) with $\Psi_{2n}$ has a similar form
\[
	\left\langle \Psi_{2n}^{\mu_1 \ldots \mu_{2n}}(x_1)
	\Psi_2(x_2) \Psi_2(x_3) \right\rangle
	= \frac{\lambda_{\phi^2\phi^2\Psi_{2n}}}
	{|x_{12}|^{d-2}|x_{13}|^{d-2}
	|x_{23}|^{d-2}}
	\left( Z^{\mu_1} \cdots Z^{\mu_{2n}}
	- \textrm{traces} \right),	
\]
where the OPE coefficients in that case are related to the previous case by
\[
	\lambda_{\phi^2\phi^2\Psi_{2n}} = 2 \, \lambda_{\phi\phi\Psi_{2n}}.
\]
Using known forms for the conformal block $g_{\Psi_{2n}}(u, v)$ of \Eq{confblockexpansion}, one can check in any dimension that this implies
\[
	\sum_{n=0}^\infty \lambda_{\phi^2\phi^2\Psi_{2n}}^2 \,
	g_{\Psi_{2n}}(u, v)
	= 4 u^{(d-2)/2} + 4 (u/v)^{(d-2)/2},
\]

\begin{figure}
	\centering
	\begin{tabular}{cccccc}
		\includegraphics{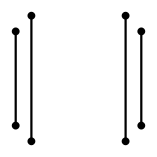} &
		\includegraphics{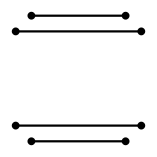} &
		\includegraphics{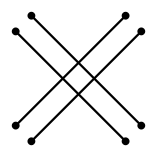} &
		\includegraphics{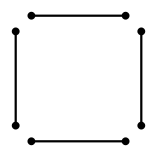} &
		\includegraphics{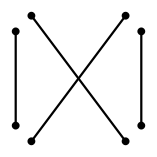} &
		\includegraphics{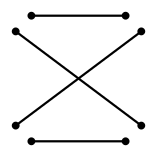} \\
		(a) & (b) & (c) & (d) & (e) & (f)
	\end{tabular}
	\caption{The 6 Feynman diagrams that contribute to the four-point function for the operator $\phi^2$. (a), (b) and (c) are disconnected, (d), (e) and (f) are connected. In terms of an OPE grouping the two points to the left and the points to the right, (a) corresponds to the exchange of the identity operator (no $\phi$), (d) and (e) to operators of type $\partial^{2n} \phi^2$, (b), (c) and (f) to operators of type $\partial^{2n} \phi^4$.}
	\label{fig:phisquared4ptfct}
\end{figure}

The remaining operators in the OPE are those constructed out of four $\phi$ fields. The problem of finding the operators and OPE coefficients in that case is more delicate, as there are not only operators of twist $\Delta - \ell = 2(d-2)$ (totally symmetric tensors), but also of higher twist. We know that their contribution in terms of conformal blocks must be
\[
	\sum\limits_{n=0}^\infty
	\lambda_{\phi^2\phi^2 \partial^{2n} \phi^4}^2 \,
	g_{\partial^{2n}\phi^4}(u, v)
	= u^{d-2} + (u/v)^{d-2} + 4(u^2/v)^{(d-2)/2}.
\]
so as to recover the complete free theory result
\[
	\eql{free4ptfct}
	& \big\langle \Psi_2(x_1) \Psi_2(x_2)
	\Psi_2(x_3) \Psi_2(x_4) \big\rangle
	\\
	& \qquad
	= \frac{1 + u^{d-2} + (u/v)^{d-2}
	+ 4 u^{(d-2)/2} + 4 (u/v)^{(d-2)/2} + 4(u^2/v)^{(d-2)/2}}
	{\left( x_{12}^2 x_{34}^2 \right)^{d-2}}.
	\nonumber
\]
The six terms in the numerator on the \rhs\ correspond to the six diagrams in \Fig{phisquared4ptfct}. The last three form the connected part of the four-point function that is being used in \Eq{freescalaranomaly}.

For completeness, we also provide the coefficient for the leading operator of type $\partial^{2n} \phi^4$ in the $\phi^2\phi^2$ OPE,
\[
	\lambda_{\phi^2 \phi^2 \partial^0\phi^4} = \sqrt{6},
	\hspace{2cm} \text{where}\quad
	\partial^0\phi^4 \equiv
	\frac{1}{2 \sqrt{6}} \, \phi^4.
\]
The operators and OPE coefficients computed in this section are used to derive the numerical results in Table~\ref{tab:scalar_d8_example}.

\subsection{The Scale Anomaly as a Cross-Section}

In the free scalar theory, $c_\scr{O}$ can easily be computed in Minkowski momentum space, providing an independent check of the result \Eq{freescalaranomaly}. We make use of its link to the imaginary part of the forward scattering amplitude of the probe field,
\[
	c_{\phi^2} = \frac{4}{\pi \, \epsilon^4} \,
	\Im \mathcal{M} (s).
	\eql{cIm}
\]
By the optical theorem, $\Im \mathcal{M}(s)$, is related to the cross-section for the scattering process $AA\to \phi\phi$, for which the amplitude can be expressed in terms of the scattering angle $\theta$ as
\[
	i\scr{M} \left( AA \to \phi \phi \right)
	= i \epsilon^2  \left[\frac{4\pi^{d/2}}{\Gamma\left(\frac{d-2}{2}\right)}\right]^2 \,
	\frac{8}{s \, \sin^2(\theta)}.
\]
Performing the phase space integral of the amplitude squared to obtain the cross-section, we have
\begin{align}
	2\Im\mathcal{M}\left( s \right)
	&= \frac{s^{d/2-2}}{2^{2d-3} \pi^{d/2-1}
	\Gamma\left( \frac{d-2}{2} \right)}
	\int_0^\pi d\theta \, \sin(\theta)^{d-3}
	\left| \scr{M} \left( AA \to \phi \phi \right) \right|^2
	\nonumber \\
	&= \epsilon ^4 \frac{2^{16-d} \pi^{3d/2 + 1}}
	{(d-6)^3 (d-4)^3 \left[\Gamma\left(\frac{d-2}{2}\right)\right]^2 \Gamma(d-6) \Gamma\left( \frac{d-6}{2} \right)} \,
	s^{d/2 - 4} .
	\eql{2ImMOPT}
\end{align}
As discussed in \sec{IRdivergences}, this rate is IR divergent in $d \leq 6$. In $d = 8$, the dependence on $s$ vanishes and we obtain
\[
	\eql{cphi2:2}
	\Im\mathcal{M}\left( s \right)
	= \frac{\epsilon^4 \pi^{13}}{16}
	\qquad
	\Rightarrow
	\qquad
	c_{\phi^2} = \frac{\pi^{12}}{4},
\]
matching the position-space result of \Eq{freescalaranomaly}.

\subsection{The Scale Anomaly in Momentum Space}
\scl{expSum}

The scale anomaly $c_{\phi^2}$ can as well be computed using the method derived in \sec{momentumspace}. We present below the details of this computation, keeping the space-time dimension $d$ arbitrary up to the end.

In momentum space, only the operators $\Psi_{2n}$ defined in \Eq{phi2operators} contribute to the anomaly, contrarily to the sum in position space that included all the operators of the $\phi^2\phi^2$ OPE. We need therefore to compute both the two-point function $\Pi_{\Psi_{2n}}(p_1 + p_2)$ defined in \Eq{Piintegral} and the three-point function $V_{\Psi_{2n}}^I (p_1,p_2)$ of \Eq{Vintegral} for the operator $\Psi_{2n}$.

We begin with the two-point function. It can be written as
\[
	\Pi_{\Psi_{2n}}(k) \delta^{IJ} =
	\epsilon^I_{\mu_1 \ldots \mu_{2n}}(k)
	\epsilon^J_{\nu_1 \ldots \nu_{2n}}(k)
	\left( -k^2 \right)^{d/2 - 2 + 2n} a_\Pi^{(n)}
	\eta^{\mu_1 \nu_1} \cdots \eta^{\mu_{2n} \nu_{2n}}.
	\eql{2ptntensor}
\]
All other possible terms that include traces like $\eta_{\mu_i\mu_j}$ or momenta $k_{\mu_i}$ vanish when contracted with the polarization tensors (remember that $\Psi_{2n}$ is a conserved current), and permutations of the indices is already taken into account from the symmetry of the polarization tensors.
The coefficients $a_\Pi^{(n)}$ can be determined in a tedious calculation to be
\begin{align}
	a_\Pi^{(n)} = & ~ \frac{1}
	{2^{2n + 4} \pi ^{(d-1)/2}}
	\frac{{\Ga \left( \frac{d-3}{2} + 2n \right)
	\Ga \left( \frac{d-2}{2} \right)}}
	{\Ga \left( d - 3 + 2n \right)}
	\nonumber \\
 	& \times \sum_{s,t = 0}^n b^{(n)}_{n-s} b^{(n)}_{n-t}
 	\sum_{r = 0}^{2s + 2t}
 	\frac{(-2)^r (2s + 2t)! \, \Ga \left( \frac{d}{2} - 1 + r \right)}
 	{r! (2s + 2t - r)! \, \Ga \left( d - 2 + r \right)},
 	\eql{aPin}
\end{align}
where the $b^{(n)}_s$ coefficients are linear combinations of the $c^{(n)}_{s,k}$ coefficients of \Eq{doubletracecoeffs},
\[
	b^{(n)}_s \equiv {2^s}\sum\limits_{k = 0}^{2n - 2s} (-1)^{s+k} c^{(2n)}_{s,k}
	= \frac{(-1)^s (2n)! \, \Gamma\left( \frac{d-2}{2} \right)
	\Gamma\left( d - 3 + 4n - 2s \right) c^{(2n)}_{0,0}}
	{s! \, (2n - 2s)! \, \Gamma\left( d - 3 + 2n \right)
	\Gamma\left( \frac{d}{2} - 1 + 2n - s \right)}.
	\eql{bns}
\]

The three point function $V_{\Psi_{2n}}^I$ must have precisely the same tensor structure as the operator $\Psi_{2n}^{\mu_1 \ldots \mu_{2n}}$ itself, with momenta $p_1$ and $p_2$ replacing the derivatives, namely
\begin{align}
	\lambda_{\phi^2\phi^2 \Psi_{2n}} V_{\Psi_{2n}}^I(p_1,p_2)
	=~ &  a_{V,n} \, s^{d/2-3} \,
	\epsilon^I_{\mu_1 \ldots \mu_{2n}}
	\sum_{s=0}^{n}
	\sum_{k=0}^{2n - 2s} (-1)^n c^{(2n)}_{s,k} \,
	\eta^{\mu_1 \mu_2} \cdots
	\eta^{\mu_{2s-1} \mu_{2s}}
	\nonumber \\
	& \quad
	\times (p_1)_{\mu_{2s+1}} \cdots (p_1)_{\mu_{2s+k}}
	(p_2)_{\mu_{2s+k+1}} \cdots (p_2)_{\mu_{2n}}
	(p_1 \cdot p_2)^s .
\end{align}
The unknown overall coefficient $a_{V,n}$ can again be determined by looking specifically at a particular component. We get
\[
	a_V^{(n)} =  - \frac{i \, \pi^{(d+2)/2}}{2^{d-6+2n}
	\left[ \Gamma\left(\frac{d-2}{2}\right) \right]^2
	(2n)! \, c_{0,0}^{(2n)}}
	\sum_{s = 0}^n b^{(n)}_{n-s} (2s)!
	\sum_{k = 0}^{2s} \frac{(-2)^k
	\Gamma\left( \frac{d}{2} - 2 + k \right)}
	{k! \, (2s - k)! \, \Gamma\left( d - 3 + k \right)}.
	\eql{aYn}
\]
Then the norm square of the three-point function is given by
\begin{align}
	\sum_I \left| \lambda_{\phi^2\phi^2 \Psi_{2n}}
	V_{\Psi_{2n}}^I(p_1,p_2) \right|^2
	& = \frac{\left[ (2n)! \right]^2}
	{2^{2n}}
	b^{(n)}_0 c^{(2n)}_{0,0} \big| a_V^{(n)} \big|^2
	s^{d - 6 + 2n},
 	\eql{3ptnorm}
\end{align}
where we have used the on-shell condition $p_1^2 = p_2^2 = 0$, as well as the definition $s = -2 p_1 \cdot p_2$.

Combining the above results for the two- and three-point functions, we obtain finally
\[
	\Im\scr{M}(s) = \frac{1}{2} \epsilon^4 s^{d/2-4} \sum_{n=0}^\infty
	\frac{\left[ (2n)! \right]^2 b^{(n)}_0 c^{(2n)}_{0,0} \big| a_V^{(n)} \big|^2}
	{2^{2n} a_\Pi^{(n)}}.
\]
In $d=8$, we recover the result of \Eq{cphi2:2},
\[
	\Im\scr{M}(s) =
	\frac{\epsilon^4 \pi^{13}}{16}
	\sum_{n=0}^\infty \frac{4n+5}{(n+1) (n+2) (2n+1) (2n+3)}
	= \frac{\epsilon^4 \pi^{13}}{16}.
\]

\newpage
\bibliographystyle{utphys}
\bibliography{mycites}

\end{document}